\documentclass[twocolumn,trackchanges]{aastex701}
\usepackage[colorinlistoftodos]{todonotes}
\usepackage{amsmath}
\newcommand{\be}{\begin{eqnarray}}
\newcommand{\ee}{\end{eqnarray}}

\usepackage{CJKutf8}
\usepackage{hyperref}
\usepackage{xcolor}

\begin{document}

\title{Numerical Non-Adiabatic Tidal Calculations with \texttt{GYRE-tides}: The WASP-12 Test Case}

\begin{CJK*}{UTF8}{gbsn}


\author[0000-0001-9037-6180,gname=Sun,sname=Meng]{Meng Sun(孙萌)}

\affiliation{National Astronomical Observatories, Chinese Academy of Sciences, 20A Datun Road, Chaoyang District, Bejing 100101, China}
\affiliation{Center for Interdisciplinary Exploration and Research in Astrophysics (CIERA), Northwestern University, 1800 Sherman Ave, Evanston, IL 60201, USA}
\email[show]{sunmeng@nao.cas.cn}

\author[0000-0002-2522-8605]{R. H. D. Townsend}
\affiliation{Department of Astronomy, University of Wisconsin-Madison, 475 N Charter St, Madison, WI 53706, USA}
\email{townsend@astro.wisc.edu}

\author[0009-0002-2514-9584, gname=Hongbo, sname=Xia]{Hongbo Xia(夏宏博)}
\affiliation{Center for Interdisciplinary Exploration and Research in Astrophysics (CIERA), Northwestern University, 1800 Sherman Ave, Evanston, IL 60201, USA}
\email{hongboxia2027@u.northwestern.edu}

\author[0000-0002-2874-2706,gname=Jifeng,sname=Liu]{Jifeng Liu}
\affiliation{National Astronomical Observatories, Chinese Academy of Sciences, 20A Datun Road, Chaoyang District, Bejing 100101, China}
\affiliation{School of Astronomy and Space Science, University of Chinese Academy of Sciences, Beijing 100049, China}
\affiliation{Institute for Frontiers in Astronomy and Astrophysics, Beijing Normal University, Beijing 102206, China}
\affiliation{New Cornerstone Science Laboratory, National Astronomical Observatories, Chinese Academy of Sciences, Beijing 100012, China}
\email{jfliu@nao.cas.cn}

\begin{abstract}

We revisit the tidal evolution of the WASP-12 system using direct numerical calculations with the \texttt{GYRE-tides} code. WASP-12b is a hot Jupiter on a 1.1-day orbit around a slightly evolved F-type star. Its observed orbital decay rate, $|\dot{P}_{\rm orb}/P_{\rm orb}| \approx 3.2\,\mathrm{Myr}^{-1}$, provides a strong constraint on stellar tidal dissipation. We confirm that linear tides with radiative damping and convective damping, as currently implemented, are not sufficient to reproduce the observed inspiral timescale. Nevertheless, our calculations, based on fully non-adiabatic forced oscillations in \texttt{MESA} stellar models with convective envelopes, yield dissipation rates that are consistent with previous semi-analytic and adiabatic estimates, confirming the robustness of our numerical framework. As the only open-source, actively maintained tool capable of computing orbital evolution in exoplanet systems, \texttt{GYRE-tides} provides a benchmark calculation for WASP-12 and future applications. {Our results validate \texttt{GYRE-tides} as a tool for analyzing combined radiative and convective damping, and indicate that the observed decay rate requires tidal dissipation operating in or near the fully damped regime, which may be achieved through nonlinear damping. These} contributions could also be evaluated by computing the wave luminosity at the radiative–convective boundary using our tool. \texttt{GYRE-tides} offers an open-source framework for computing tidal dissipation in short-period exoplanet systems, including the many systems expected to show orbital decay in upcoming Roman surveys.

\end{abstract}


\section{Introduction\label{sec:intro}}

The tidal evolution of short-period hot Jupiter systems provides critical insights into the interplay between stellar structure, fluid dynamics, and orbital dynamics. Among these systems, WASP-12 stands out as a unique laboratory for studying tidal dissipation due to its remarkably rapid orbital decay. Several previous studies have investigated whether the orbital decay of WASP-12b, characterized by a period derivative of $\dot{P}_{\rm orb}/P_{\rm orb} \approx -3.2\,\mathrm{Myr}^{-1}$, can be accounted for by stellar tidal dissipation. \citet{Chernov2017,Weinberg2017,Barker2020} explored whether nonlinear wave breaking of the dynamical tide near the stellar center could provide the necessary dissipation. They concluded that, for main-sequence models with a convective core, radiative damping acting on dynamical tides and convective damping acting on equilibrium tides near the surface are both too inefficient. However, in post-main-sequence subgiant models with radiative cores, nonlinear breaking of inward-propagating, tidally excited gravity waves becomes viable, leading to enhanced dissipation consistent with the observed decay rate. In addition to the strongly nonlinear prescription, \citet{Weinberg2024} studied WASP-12 in the weakly nonlinear regime, where the primary mode excites a sea of secondary modes through three-mode interactions, and calculated the net mode dissipation across a range of stellar masses and evolutionary stages.

Analyzing the problem from a dynamics perspective, \citet{Millholland2018} proposed a planetary obliquity tide model, suggesting that the high obliquity of WASP-12b could be maintained through a secular spin–orbit resonance with a third planet of 10–20 Earth masses. In their model, the perturber would need to reside at a semimajor axis of less than 0.04 AU. Notably, this mechanism can reproduce the observed orbital decay rate without requiring the host star to be a subgiant. \citet{Bailey2019} re-examined the evolution of WASP-12 using a broader and denser stellar model grid and found that if WASP-12 were still on the main sequence with a convective core, the g-mode resonances would remain narrow, requiring implausibly fine-tuned orbital frequencies to explain the observed decay rate. Moreover, constructing a subgiant model that simultaneously matches all observed parameters under standard mixing and initial metallicity assumptions proves challenging.

Magnetic fields have also been proposed to enhance tidal dissipation in stellar interiors. \citet{Wei2022} demonstrated that turbulent Ohmic dissipation acting on equilibrium tides can be much more efficient than turbulent viscous dissipation. Most recently, \citet{Duguid2024} proposed another efficient tidal dissipation mechanism that requires a strong magnetic field. In this scenario, tidally excited inward-propagating internal gravity waves can convert into outward-propagating magnetic waves. These magnetic waves then dissipate in regions with lower Alfvén velocity. This process can produce a level of tidal dissipation comparable to that of {the fully damped (nonlinear wave breaking)} regime, even in stars with convective cores. It also explains the observed orbital decay rate of WASP-12b and could more broadly influence the tidal evolution of hot Jupiters and other close-in planets orbiting F-type stars.

We revisit the WASP-12 system using {linear hydrodynamical} direct numerical tidal calculations implemented in \texttt{GYRE-tides} \citep{Townsend2013,Townsend2018,Goldstein2020,Sun2023}. The study highlights the role of \texttt{GYRE-tides} as an open-source tool for reproducible calculations of tidal dissipation and secular evolution in exoplanet systems. Our goal is to provide reproducible tidal calculations with \texttt{GYRE-tides}, which recover the dissipation from radiative diffusion and convective damping in agreement with previous studies, while also enabling analyses of potential nonlinear contributions to orbital evolution. This Letter is structured as follows: Section~\ref{sec:obs} briefly summarizes the observation properties about the WASP-12 system; Section~\ref{sec:results} introduces the stellar model and presents detailed numerical results on the tidal secular evolution of WASP-12b; Section~\ref{sec:discussions} analyzes the possible damping mechanisms for causing the orbital decay, and Section~\ref{sec:conclusions} presents our conclusions and implications for future work.

\section{Observations\label{sec:obs}}
\begin{table*}[t]
\centering
\caption{Stellar Parameters for the WASP-12 Host Star from the Literature}
\label{tab:stellar_params_lit}
\begin{tabular}{ccccccc}
\hline
$T_{\mathrm{eff}}$ [K] & [Fe/H] & $\bar{\rho}$ [g\,cm$^{-3}$] & $M_\star$ [$M_\odot$] & $R_\star$ [$R_\odot$] & $\log_{10}(g\,\mathrm{[cm\,s^{-2}]})$ & Reference \\
\hline
$6265 \pm 50$       & $0.12 \pm 0.07$ & $0.386 \pm 0.014$               & $1.325^{+0.026}_{-0.018}$ & $1.690^{+0.019}_{-0.018}$ & $4.11 \pm 0.11$  & \citet{Leonardi2024} \\
$6154 ^{+106}_{-105}$      & --- & $0.308^{+0.081}_{-0.055}$               & $1.170^{+0.184}_{-0.138}$ & $1.749^{+0.069}_{-0.094}$ & $4.02^{+0.08}_{-0.08}$ & \citet{Stassun2019} \\
$6360 \pm 140$      & ---            & ---                             & $1.434 \pm 0.110$          & $1.657 \pm 0.046$         & ---              & \citet{Chakrabarty2019} \\
$6300^{+200}_{-100}$ & $0.30 \pm 0.10$ & $0.492 \pm 0.084$       & $1.35 \pm 0.14$            & $1.57 \pm 0.07$           & $4.17 \pm 0.03$  & \citet{Ozturk2019} \\
$6269^{+604}_{-408}$ & ---            & ---                             & ---                        & $1.571^{+0.227}_{-0.264}$ & ---              & \citet{Gaia2018} \\
$6250 \pm 100$      & $0.32 \pm 0.12$ & ---                             & $1.38 \pm 0.18$            & $1.619 \pm 0.076$         & ---              & \citet{Bonomo2017} \\
$6300 \pm 150$      & $0.30$          & $0.42 \pm 0.07$                 & $1.20 \pm 0.45$            & $1.59 \pm 0.18$           & $4.38 \pm 0.10$  & \citet{Stassun2017} \\
$6360^{+130}_{-140}$    & $0.33^{+0.14}_{-0.17}$ & $0.446^{+0.015}_{-0.014}$       & $1.434^{+0.110}_{-0.090}$  & $1.657^{+0.046}_{-0.044}$ & $4.157^{+0.013}_{-0.012}$ & \citet{Collins2017} \\
$6300$                 & $0.30$                 & ---                             & ---                        & ---                       & $4.38$                 & \citet{Turner2016} \\
---                    & $0.070 \pm 0.070$      & ---                             & $1.38 \pm 0.19$            & ---                       & ---                    & \citet{Knutson2014} \\
$6313 \pm 52$           & $0.21 \pm 0.04$        & $0.22 \pm 0.02$                 & $1.26 \pm 0.10$            & $1.21 \pm 0.21$           & $4.37 \pm 0.12$         & \citet{Mortier2013} \\
$6250 \pm 100$          & $0.32 \pm 0.12$        & $0.458 \pm 0.023$               & $1.38 \pm 0.19$            & $1.619 \pm 0.079$         & $4.159 \pm 0.024$       & \citet{Southworth2012} \\
$6300 \pm 200$          & ---                    & ---                             & $1.35 \pm 0.14$            & $1.57 \pm 0.07$           & ---                    & \citet{Hebb2009} \\
\hline
\end{tabular}
\end{table*}

The WASP-12 system has been the subject of sustained observational interest over the past decade due to mounting evidence that its short-period hot Jupiter is undergoing measurable orbital decay. Numerous studies have reported the host star's effective temperature to lie in the range of $T_{\mathrm{eff}} \sim 6050\,\text{--}\,6500\,\mathrm{K}$, with a slight metallicity enrichment relative to the Sun, typically $[\mathrm{Fe}/\mathrm{H}] \sim 0.1\,\text{--}\,0.3$. The mean stellar density spans a relatively broad range of $\bar{\rho} \sim 0.2\,\text{--}\,0.5\,\mathrm{g\,cm^{-3}}$. Reported stellar masses and radii fall in the intervals $M_\star \sim 1.1\,\text{--}\,1.4\,M_\odot$ and $R_\star \sim 1.5\,\text{--}\,1.8\,R_\odot$, respectively, while surface gravities are generally consistent with $\log g \sim 4.0\,\text{--}\,4.4\,\mathrm{cm\,s^{-2}}$.

The transiting planet WASP-12b is a typical hot Jupiter in terms of mass, with $M_p$ estimates ranging from 1.3 to 1.5 $M_{\mathrm{Jup}}$ across different studies. Most measurements converge around approximately 1.4 $M_{\mathrm{Jup}}$. The planet orbits extremely close to the host star on a nearly circular orbit. Its orbital period is measured with high precision to be 1.0914 days, and the system exhibits orbital decay on a timescale of a few million years. Notably, the host star's projected rotational velocity is very low, suggesting slow stellar rotation that is not synchronized with the orbit. The close separation and short decay timescale have motivated continuous photometric and spectroscopic monitoring campaigns aimed at confirming the long-term decrease in the orbital period \citep[e.g.,][]{Maciejewski2016,Patra2017,Yee2020}.

We compile the key observed stellar properties in Table~\ref{tab:stellar_params_lit}, including those relevant for \texttt{MESA} modeling such as $T_{\mathrm{eff}}$, $\bar{\rho}$, $M_{\star}$, and $R_{\star}$, for reference and comparison.

\section{Calculations\label{sec:results}}
We present our numerical analysis of tidal secular evolution for the WASP-12 system, using the direct numerical approach implemented in \texttt{GYRE-tides}. Our goal is to assess whether linear tidal dissipation mechanisms, including radiative damping and convective damping due to turbulent eddies, or additional damping mechanisms, are required to reproduce the observed orbital decay rate of WASP-12b to the correct order of magnitude.

Stellar models are constructed using \texttt{MESA} version r24.08.1 \citep{Paxton2011,Paxton2013,Paxton2015,Paxton2018,Paxton2019,Jermyn2023}. {Our inlists and analysis scripts are available on Zenodo \dataset[10.5281/zenodo.17532140]{https://doi.org/10.5281/zenodo.17532140}.} The fiducial model represents a $1.3\,M_\odot$ main-sequence star with a metal-rich metallicity setting of $Z = 0.025$, a radius of $1.55\,R_\odot$, an effective temperature of $T_{\rm eff} = 6216\,\mathrm{K}$, and an average density of $\bar{\rho} \sim 0.49\,\mathrm{g\,cm^{-3}}$. We adopt a mixing length parameter of $\alpha = 2.0$. This model is not intended to precisely reproduce the observed stellar parameters but serves as a reference for the detailed tidal response analysis presented in the following sections. 

We adopt the \texttt{MIST} core overshooting prescription, which has been calibrated using the main-sequence turnoffs of open cluster M67 \citep{Choi2016}. Specifically, we apply exponential overshooting at the convective core boundary during hydrogen burning, with parameters \texttt{overshoot\_f = 0.016}, \texttt{overshoot\_f0 = 0.008}, and \texttt{D\_min = 0.01}. These settings smooth composition gradients and yield physically consistent wave-like responses.

For the orbital configuration, we adopt a planet mass of $1.4\,M_{\rm Jup}$ (treated as a point mass), an eccentricity of $0.01$, and a primary stellar rotation rate set to 1\% of the orbital frequency at periastron, corresponding to slow rotation. {We note that our code requires a small but finite $e$ for numerical stability because some secular coefficients scale as $1/e$; tests with $e=10^{-3}$ (and smaller) yield indistinguishable dissipation and orbital decay rates. We assume spin orbit alignment and solid body stellar rotation; rotation affects the response only through the forcing frequency, and we neglect Coriolis modifications to the mode spectrum.} 
While treating WASP-12b as a point mass is a simplification, detailed studies have revealed that this planet is significantly distorted by tidal forces, with a prolate shape and an extended atmosphere that likely overflows its Roche lobe \citep{Li2010,Lai2010}. We have also tested alternative stellar models, including ones without core overshooting and others that better match the observed stellar and orbital properties. All cases yield similar orbital decay timescales, indicating that our tidal dissipation results are not particularly sensitive to the stellar structure within the plausible range for the WASP-12 system. Tides raised on the planet are neglected in this work. 

\begin{figure}[t]
    \centering
    \includegraphics[width=1\linewidth]{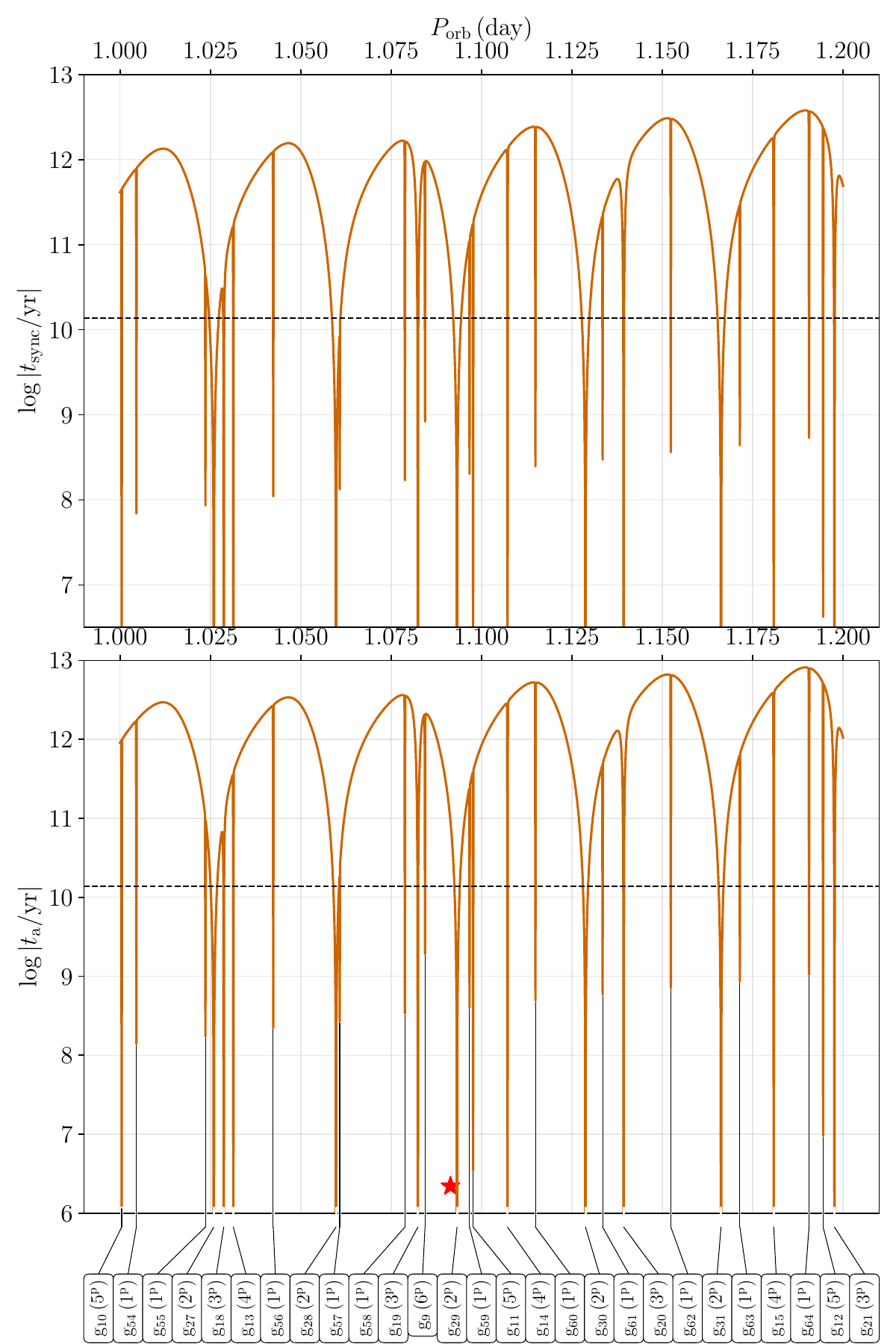}
    \caption{
    Synchronization timescale $t_{\rm sync}$ (top panel) and orbital decay timescale $t_a$ (bottom panel) as functions of orbital period, computed for the WASP-12 system considering only radiative damping of the dynamical tide. The results are based on linear tidal response functions calculated with \texttt{GYRE-tides}. Sharp decreases in the timescales arise from resonance with g-modes; the labels beneath the plot identify the resonant mode, and additionally indicate (in parentheses) the orbital harmonic $k$ involved in the resonance, and the propagation direction (``p'': prograde; ``r'': retrograde) of the mode. {The horizontal dashed lines in both panels denote the age of the Universe}. Outside the resonances, the predicted timescales generally exceed the observed decay timescale by approximately three orders of magnitude, indicating that radiative damping alone cannot explain the rapid inspiral of WASP-12b. The red star marks the observed value of the orbital decay timescale, $|t_a| \sim 2.2\,\mathrm{Myr}$.
    }
    \label{fig:fig_secular_0.01pseudosync_rad_only}
\end{figure}

We first examine the case in which only radiative damping of dynamical tides is included (Figure~\ref{fig:fig_secular_0.01pseudosync_rad_only}). In this scenario, the predicted orbital decay timescale ($t_a$) remains several orders of magnitude longer than the observed value for WASP-12b, typically ranging from approximately $10^{12}$ to $10^{13}$ years across most orbital periods. Significant drops in $t_a$ occur only at a few narrow resonances, such as those corresponding to high-order g-modes like $\mathrm{g}_{27}$ and $\mathrm{g}_{28}$.  These dips reflect strong tidal coupling at specific frequencies, where dynamical tides can transfer angular momentum more efficiently. However, outside these resonance regions, the timescales are still too long to explain the observed orbital decay rate of $|\dot{P}_{\rm orb}/P_{\rm orb}| \approx 3.2\,\mathrm{Myr}^{-1}$. These results are roughly consistent with the findings of \citet{Chernov2017}, \citet{Weinberg2017}, and \citet{Bailey2019}.

The synchronization timescale ($t_{\rm sync}$) shows a similar trend to $t_a$ as a function of orbital period. The absolute value of $t_{\rm sync}$ is shorter than $t_a$ by less than an order of magnitude. This offset can be understood by comparing the stellar spin angular momentum $J_\star$ to the orbital angular momentum $J_{\rm orb}$. For a $1.3\,M_\odot$ star with a radius of $1.6\,R_\odot$, a moment of inertia constant of 0.05, and a surface rotation speed of 2 km/s, and assuming a hot Jupiter companion with a mass of $1.4\,M_{\rm Jup}$, we estimate the ratio $J_\star / J_{\rm orb}$ to be approximately 0.2. This value is consistent with the relative difference between $t_{\rm sync}$ and $t_a$. Figure \ref{fig:fig_secular_0.01pseudosync_rad_only} confirms that in the absence of additional dissipation mechanisms, linear dynamical tides alone are insufficient to reproduce the observed orbital decay rate. {However, if the system were fortunate enough to reside in a resonance, the dense forest of $g$-mode resonances naturally present in F-type stars could enhance the tidal dissipation.}

{A related explanation is resonance locking, where the forcing and mode frequencies co-evolve to stay near resonance (see \citealt{Ma2021} for a modal decomposition method). By contrast, our linear \texttt{GYRE-tides} framework uses a direct-solution approach, solving the forced oscillation equations and boundary conditions without normal-mode expansion. In our WASP-12 models, the dense $g$-mode spectrum admits narrow resonances that, under slight shifts in stellar parameters, could allow temporary locking; however, such locking could be unstable once wave breaking becomes important, especially after the convective core recedes and the core becomes fully radiative. A full quantitative treatment is left for future work.}

\begin{figure}[t]
    \centering
    \includegraphics[width=1\linewidth]{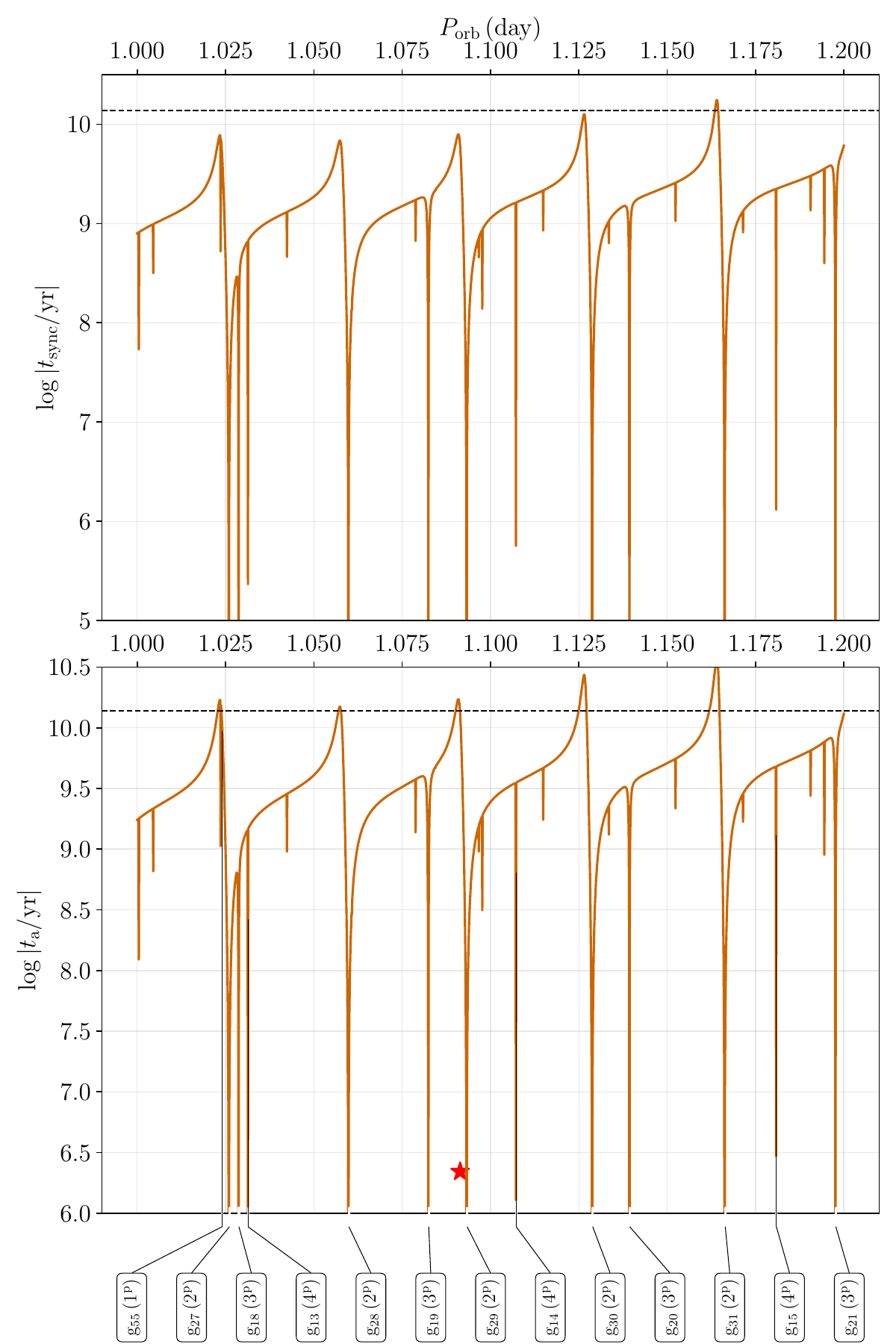}
    \caption{
As in Figure~\ref{fig:fig_secular_0.01pseudosync_rad_only}, the same binary configuration and stellar structure are used, but here both with convective damping and radiative damping are included. The inclusion of convective damping enhances tidal dissipation across a broader range of orbital periods. The red star denotes the observed decay timescale.
}
\label{fig:fig_secular_0.01pseudosync}
\end{figure}

When convective damping of equilibrium tides is included (Figure~\ref{fig:fig_secular_0.01pseudosync}), the secular evolution timescales are dramatically reduced. {In our full numerical solution, convective damping acts on both the equilibrium tide and the evanescent tails of the internal gravity waves that penetrate the convective envelope, as the linear tidal response is obtained by solving an inhomogeneous boundary value problem.} The upper envelope of the synchronization timescale decreases to approximately $10^{9}$–$10^{10}$ years, while the orbital decay timescale drops to around $10^{9.5}$–$10^{10.5}$ years near the current observed orbital period of $\sim 1.09$ days. Although some high-order $g$-modes are resonantly excited, they {may} correspond to brief sweep-by events during the orbital evolution. Therefore, the overall envelope of the timescales remains markedly longer than the observed value. This indicates that equilibrium tides, even when efficiently damped by turbulent convection in the stellar envelope, are insufficient to account for the rapid orbital decay of WASP-12b.

\section{Analysis\label{sec:discussions}}
The goal of this discussion is to evaluate the impact of different tidal dissipation mechanisms on the WASP-12 system. We first present the internal structure of a fiducial stellar model, and then analyze the roles of radiative diffusion, convective damping, and potential nonlinear effects in driving the orbital evolution.

\begin{figure}[t]
    \centering
    \includegraphics[width=0.96\linewidth]{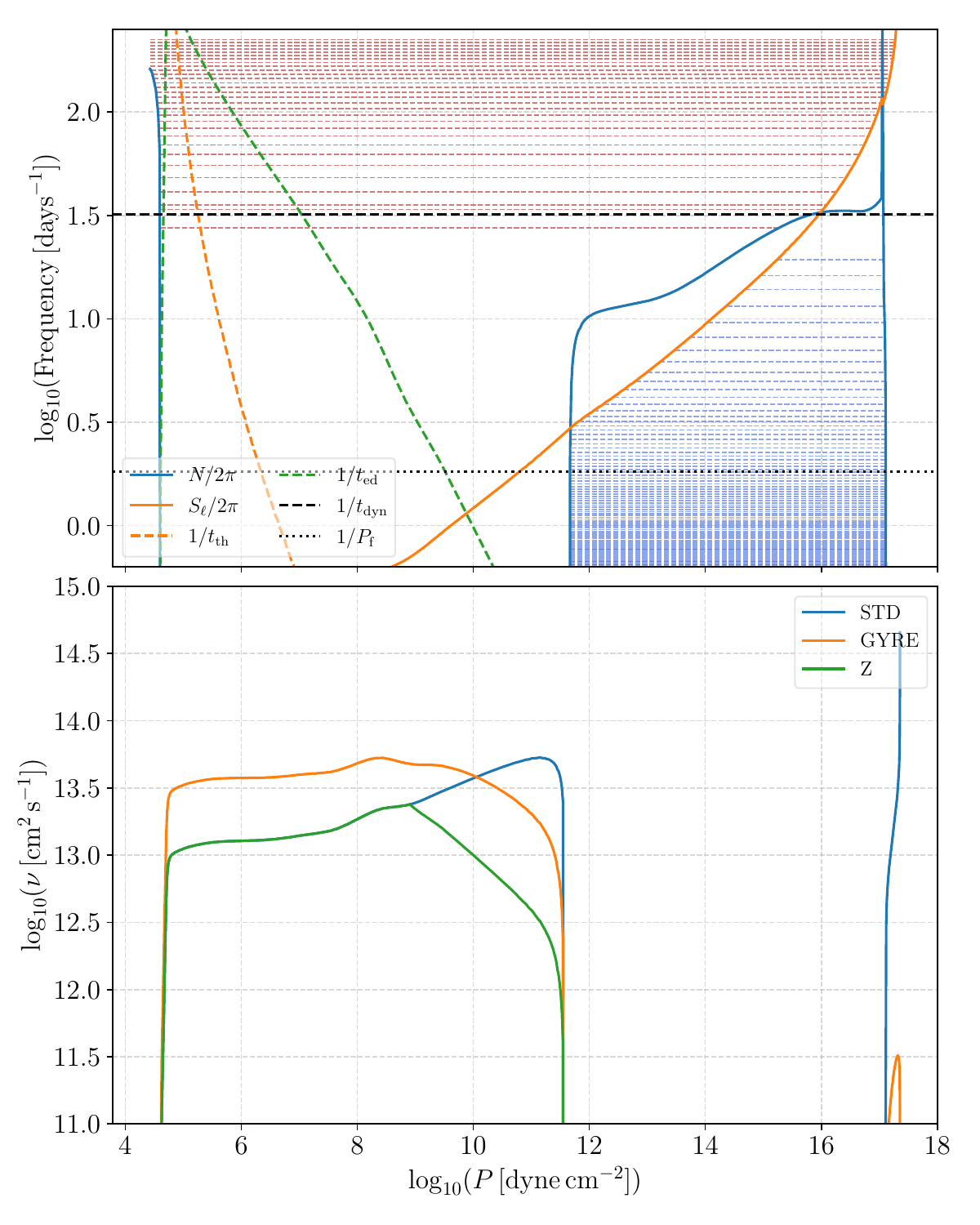}
    \caption{
    Characteristic frequencies, viscosity prescriptions, radial displacement amplitudes, and the gravitational potential perturbation response for a $1.3\,M_\odot$ main-sequence stellar model, as a function of pressure. 
    Top: The plotted quantities include the Brunt–Väisälä frequency $N/2\pi$ (blue), the Lamb frequency $S_2/2\pi$ for $\ell=2$ (orange), the inverse thermal timescale $1/t_{\rm th}$ (orange dashed line), the inverse eddy turnover timescale $1/t_{\rm ed}$ (green dashed line), the dynamical frequency $1/t_{\rm dyn}$ (horizontal black dashed line), and the tidal forcing frequency $1/P_f$ (horizontal black dotted line). Horizontal royal blue and coral lines mark the propagation regions of multiple $g$-modes and $p$-modes, respectively. 
    Bottom: Viscosity profiles standard mixing-length formulation (${\rm STD}$, blue solid line), eddy viscosity implemented in GYRE (orange solid line), and Zahn's reduced viscosity (${\rm Z}$, green solid line). 
    }
\label{fig:profile_analysis}
\end{figure}

The top panel of Figure~\ref{fig:profile_analysis} shows the propagation diagram for our fiducial 1.3 $M_\odot$ main-sequence stellar model as a function of $\log_{10}(P\,\mathrm{[dyn/cm^2]})$, where $P$ denotes the gas pressure. The plot includes the Brunt–Väisälä (buoyancy) frequency $N$, the Lamb frequency $S_\ell$ for $\ell = 2$ modes, and the inverse thermal, eddy turnover, and dynamical timescales ($1/t_{\mathrm{th}}$, $1/t_{\rm ed}$, and $1/t_{\mathrm{dyn}}$). {Horizontal black dashed and dotted lines indicate the dynamical frequency \(1/t_{\rm dyn}\) and the tidal forcing frequency \(1/P_{\rm f}\), respectively, where \(P_{\rm f} \equiv 2\pi / \lvert k\,\Omega_{\rm orb} - m\,\Omega_{\rm rot} \rvert\). For WASP-12, \(\Omega_{\rm rot} \ll \Omega_{\rm orb}\); adopting the dominant quadrupole component \((k=m=2)\) gives \(P_{\rm f} \simeq P_{\rm orb}/2\).} Very roughly, internal gravity modes reside below the $1/t_{\mathrm{dyn}}$ in frequency space, while acoustic waves lie above it. The structure of these characteristic frequencies reveals the mode propagation cavities. These propagation conditions apply under the WKB and Cowling approximations \citep{Cowling1941,Unno1989}, where the mode behavior is governed by the local sign of the squared radial wavenumber. Internal gravity waves can propagate in regions where both $N^2 > \omega_{\rm mode}^2$ and $S_\ell^2 > \omega_{\rm mode}^2$, whereas acoustic waves propagate where $N^2 < \omega_{\rm mode}^2$ and $S_\ell^2 < \omega_{\rm mode}^2$. Here $\omega_{\rm mode}$ is the mode frequency. These conditions determine the spatial domains within the star where different classes of stellar oscillation modes can exist and interact with tidal forcing.

\subsection{Radiative Diffusion Damping}

Radiative diffusion damping occurs when a mode’s period exceeds the local thermal diffusion timescale, in which case the mode is efficiently damped by radiative energy transport. This effect is important for high-order g-modes, which have long periods and are more susceptible to thermal damping. In the upper panel of Figure~\ref{fig:profile_analysis}, this condition corresponds to regions where the $1/t_{\mathrm{th}}$ curve (orange dashed line) penetrates the g-mode cavity. In our fiducial model, however, the orbital frequency (and harmonics thereof) remains well above the $1/t_{\mathrm{th}}$ line over most of the g-mode propagation region, indicating that most g-modes are not significantly affected by radiative damping. As a result, they retain their oscillatory wave structure but play a negligible role in secular tidal evolution owing to weak radiative damping (Figure~\ref{fig:fig_secular_0.01pseudosync_rad_only}). Although radiative diffusion damping renders most g-modes ineffective for long-term orbital decay, the secular profiles exhibit further structure arising from tidal harmonics. In addition to the sharp dips, a broader envelope is evident in the background of both the $t_{\rm sync}$ and $t_a$ profiles. {We interpret this envelope as being set primarily by the dominant $k=m=2$ component, whose coupling spans a wider frequency range and therefore produces an extended depression in the orbital decay timescale.}

With \texttt{GYRE-tides}, we can compute the tidal orbital decay produced solely by radiative diffusion by solving the non-adiabatic, forced oscillation equations. From Figure~\ref{fig:fig_secular_0.01pseudosync_rad_only}, the inferred semimajor-axis decay timescale is $t_a\sim10^{12}$–$10^{13}\,{\rm yr}$ (neglecting the resonance dips). The corresponding orbital energy dissipation rate follows from $|\dot E_{\rm orb}| = G\,(M_\star M_p)/(2a t_a)$. For $M_\star=1.3\,M_\odot$, $M_p=1.4\,M_{\rm Jup}$, and $P_{\rm orb}=1.09$ day, we obtain $|\dot E_{\rm orb}|\simeq 10^{24}$–$10^{25}$ ${\rm erg\,s^{-1}}$ for WASP-12.

\subsection{Convective Damping}

The lower panel of Figure~\ref{fig:profile_analysis} gives the turbulent viscosity profiles for three representative prescriptions: standard, Zahn-reduced, and the \texttt{GYRE-tides} implementation. The standard model adopts an unreduced mixing-length-based prescription \citep{Verbunt1995}:
\begin{equation}
\nu_{\rm STD} = \frac{1}{3} v_{\rm ed} \alpha H,
\end{equation}
where $v_{\rm ed}$ {is the mixing-length theory convective velocity from the \texttt{MESA} model}, $\alpha=2.0$ is the mixing-length parameter used for our default model, and $H$ is the local pressure scale height.

The Zahn-reduced prescription \citep{Zahn1989} introduces a frequency-dependent linear suppression factor when the eddy turnover time $t_{\rm ed}=\alpha H/v_{\rm ed}$ exceeds the tidal forcing period $P_{\rm f}/2$:
\begin{equation}
\nu_{\rm Z} = \nu_{\rm std} \times \min\left(1, \frac{P_{\rm f}}{2t_{\rm ed}}\right).
\end{equation}

The \texttt{GYRE-tides} implementation of turbulent damping follows \citet{Willems2010}, specifically their Equation~14 in Section~3 (originally from Equation 12 of \citealt{Terquem1998}, but only for the radial component), in which a radial viscous force is added to the right-hand side of the momentum equation. The effective viscosity (Equation~68 of \citealt{Willems2010}) is given by:
\begin{equation}
\nu_{\rm GYRE} = \frac{L^2}{t_{\rm ed}} \left[1 + \left( \frac{t_{\rm ed} \sigma_{m,k}}{2\pi} \right)^s \right]^{-1},
\label{nu_gyre}
\end{equation}
where $L=\alpha H$ is the mixing length, $\sigma_{m,k} = k\Omega_{\rm orb} - m\Omega_{\rm rot}$ is the tidal forcing frequency, and $s$ is typically set to $1$ (a slightly different approach than Zahn's prescription). Here, $\Omega_{\rm orb}=2\pi/P_{\rm orb}$ is the mean orbital angular frequency of the companion, $\Omega_{\rm rot}$ to be the convective envelope rotation rate. {We assume the star rotates rigidly with this single angular velocity $\Omega_{\rm rot}$, neglecting differential rotation and core envelope shear.} $m$ is the azimuthal wavenumber, and $k$ is the Fourier index associated with the tidal forcing frequency decomposition (i.e., orbital harmonics).

As shown in panel~2 of Figure~\ref{fig:profile_analysis}, the standard and Zahn prescriptions yield the same magnitudes across most of the surface convection zone, while the Zahn-reduced viscosity drops off significantly near the radiative–convective boundary where $\tau_{\rm ed} \gg 2P_{\rm f}$. For $\nu_{\rm GYRE}$, the values remain roughly half a dex higher than $\nu_{\rm STD}$ up to $\log P \sim 9$, after which they begin to be reduced. In the core region, where $\tau_{\rm ed} \gg 2P_{\rm f}$ happens again, both the Zahn and \texttt{GYRE} viscosities are significantly reduced. These differences in turbulent viscosity lead to substantial variations in the resulting tidal dissipation efficiency.

By incorporating Willems' form of radial viscous force into our \texttt{GYRE-tides} framework, we find that the orbital decay timescale $t_a$ for WASP-12b can not reach the observed value of $|\dot{P}_{\rm orb}/P_{\rm orb}| \sim \mathrm{Myr}^{-1}$. To quantify the energy dissipation rate due to turbulence, we computed the integral following Equation 5 of \citet{Sun2018}, across the entire star. This calculation requires the radial and horizontal displacement components $\xi_r$ and $\xi_h$, which we obtain from \texttt{GYRE}'s forced tidal response for the $\ell = m = k = 2$ component; the horizontal-shear terms are dropped because only the radial viscous force is included here. For our fiducial model, the resulting energy dissipation rate is 
$5.7\times10^{28}\,\mathrm{erg\,s^{-1}}$ (standard prescription), 
$1.3\times10^{26}\,\mathrm{erg\,s^{-1}}$ (Zahn’s prescription), 
and $7.4\times10^{26}\,\mathrm{erg\,s^{-1}}$ (\texttt{GYRE} viscosity prescription). Using the analytical adiabatic equilibrium tide amplitude, $\xi_{r,\rm eq}=-U_{\ell m}(r)/g$, where $U_{\ell m}(r)$ is the tidal potential and $g$ is the gravity, in the viscous integral above gives a $\simeq 10^{27}\,\mathrm{erg\,s^{-1}}$ dissipation rate. The corresponding orbital decay timescale is $\simeq10^{10}$ yr, which remains far longer than the observed Myr-level inspiral time of WASP-12b, yet is consistent with the estimates of \citet{Weinberg2017,Bailey2019}. {In addition, \citet{Terquem1998} derived an irrotational equilibrium tide, and the resulting convective damping rates typically differ by tens of percent.}

As an independent check, we also compute the tidal power using our own harmonic-sum formulation implemented in \texttt{GYRE-tides}. This begins from the local energy deposition rate, considering a fluid element of volume $\mathrm{d} \tau$. The instantaneous energy deposition rate $\dot{e}$ into this element is given by the work done by the tidal potential, namely
\begin{equation}
\mathrm{d}\dot{e} = -\rho \,\nabla \Phi_T \cdot \mathbf{v} \mathrm{d} \tau,
\end{equation}
where $\rho$ is the density, $\Phi_T$ is the tidal potential, and $\mathbf{v}$ is the velocity field. Neglecting higher-order terms and using the orthogonality relations of spherical harmonics to collapse the angular integrals, one obtains the differential (per-radius) deposition power. By integrating this quantity with respect to the radial coordinate, then averaging over an orbital period and carrying out the harmonic expansion, the expression reduces to the global harmonic-sum form.
\begin{align}
\langle P \rangle
& =  4 q^2 \frac{G M_{\star}^2}{a}
\sum_{\ell,m,k\ge 0}
\left(\frac{R_{\star}}{a}\right)^{\ell+3}
\left(\frac{r_b}{R_{\star}}\right)^{\ell+1} \nonumber \\
& \kappa_{\ell,m,k}\,
\big|\bar{F}_{\ell,m,k}\big|\,
\sin\bar{\gamma}_{\ell,m,k}\,
\bar{G}^{(5)}_{\ell,m,k},
\end{align}

and 
\begin{equation}
\bar{G}^{(5)}_{\ell,m,k}
\equiv k\,\Omega_{\rm orb}\,
\frac{2\ell+1}{4\pi}\,
\left(\frac{R_{\star}}{a}\right)^{-\ell+2}\,
\big|\bar{c}_{\ell,m,k}\big|^2,
\end{equation}
where mass ratio $q = M_{\rm p}/M_{\star}$, $r_b$ is the surface boundary radius (we adopt $r_b=R_{\star}$), $\bar{\gamma}_{\ell,m,k}=\arg(\bar{F}_{\ell,m,k})$ is the phase lag of the complex response $\bar{F}_{\ell,m,k}$, and $\kappa_{\ell,m,k}$ is the dimensionless weight of each tidal harmonic (see \citealp{Sun2023} for a definition of these quantities). The net tidal dissipation within the star is then
\[
\dot{E} \;=\; \langle P \rangle \;-\; \Omega_{\rm rot}\,\mathcal{T}_{\rm sec},
\]
with $\mathcal{T}_{\rm sec}$ the secular tidal torque.

For the fiducial model, this method yields an energy dissipation rate of \(\simeq 8.5\times10^{26}\,{\rm erg\,s^{-1}}\), demonstrating the consistency between the two evaluation methods.

{While we adopt Equation~\ref{nu_gyre} with an $s=1$ high frequency reduction, other numerical and theoretical studies indicate a stronger suppression of equilibrium tide dissipation in the fast tide regime, with an effective turbulent viscosity scaling as $(t_{\rm ed}\sigma_{m,k})^{2}$ (i.e., $s\simeq2$; \citealt{Ogilvie2012,Duguid2020b}). More realistic local hydrodynamical simulations of convection–tide coupling are presented by \citet{Penev2009,Duguid2020a,Vidal2020a,Vidal2020b,deVries2023}; see also the recent overview by \citet{Barker2025}. Our $s=1$ prescription could be regarded as an upper bound on the equilibrium-tide dissipation in WASP-12; adopting a steeper high-frequency suppression would further weaken the convective damping contribution.}

\subsection{Nonlinear Damping}

\begin{figure}[t]
    \centering
    \includegraphics[width=1\linewidth]{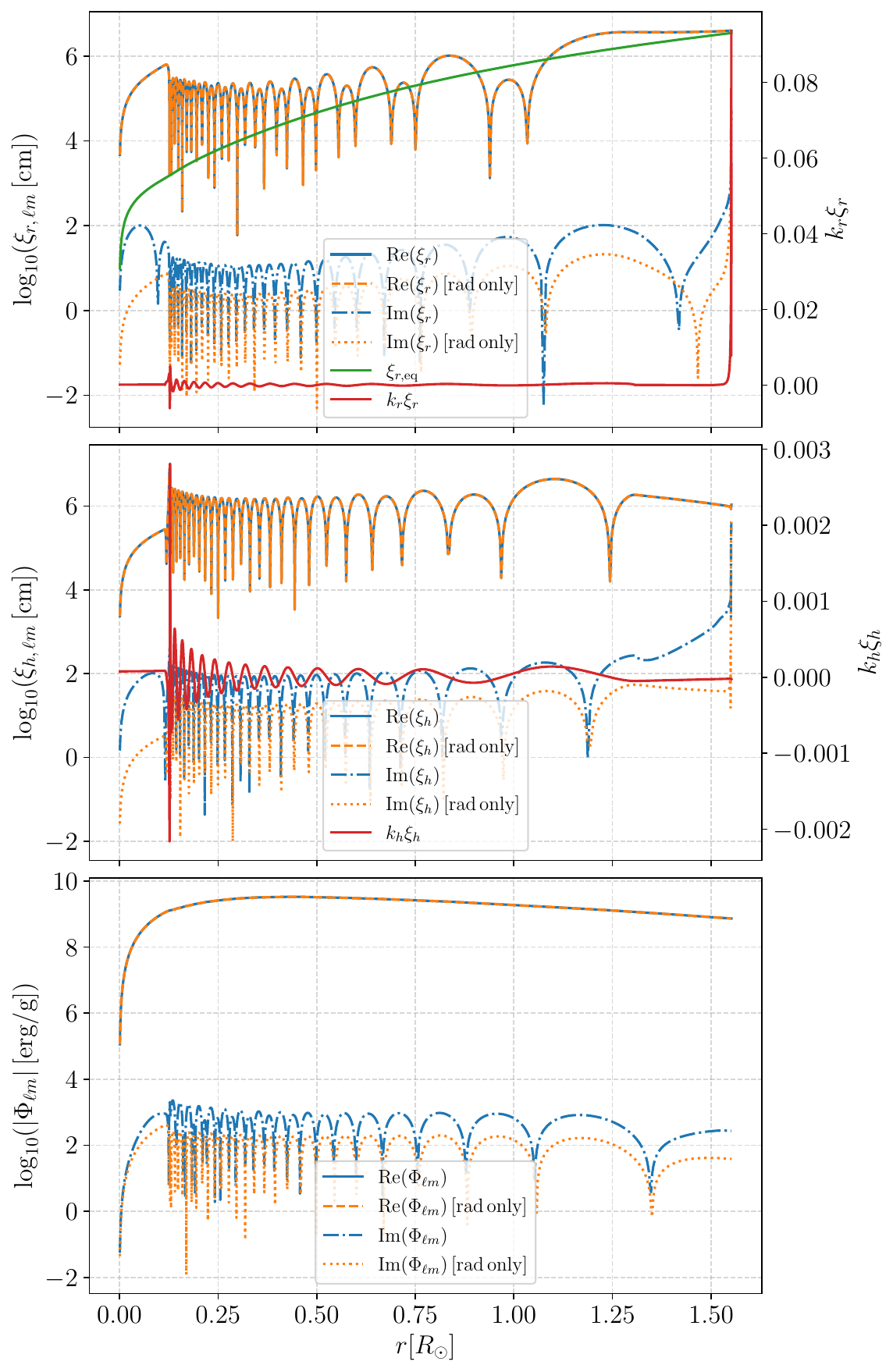}
    \caption{Mode analysis of the $\ell=m=k=2$ tidal response in our fiducial 1.3 $M_{\odot}$ main-sequence stellar model. Top: Radial displacement amplitude $\xi_r$ (real and imaginary parts), including results from the full nonadiabatic forced oscillation calculation, the case with radiative diffusion damping only, and the analytical equilibrium tide (green solid); the dimensionless nonlinearity parameter $k_r\xi_r$ is also shown as red solid line. 
    Middle: Horizontal displacement amplitude $\xi_h$ (real and imaginary parts), compared with the case with radiative diffusion damping only; the corresponding $k_h\xi_h$ is indicated as red solid line. 
    Bottom: Eulerian gravitational potential perturbation $\Phi_{\ell m}$ (real and imaginary parts), with both full and radiative-only solutions. All quantities are plotted as a function of radius.}
    \label{fig:mode_analysis_k_eq_2_ell_eq_2_MS}
\end{figure}

For the circular-orbit case, the $k=m=2$ component dominates the tidal response, as expected. In all three panels of Figure~\ref{fig:mode_analysis_k_eq_2_ell_eq_2_MS}, the imaginary parts of the displacements and Eulerian potential perturbation are much smaller than the real parts, indicating that the response is nearly adiabatic for an F-type main-sequence star, consistent with the findings of \citet{Pfahl2008}. Noticeable non-adiabatic effects arise only in the outermost layers ($r/R_\star\simeq1$), where the local thermal timescale becomes short. Some of the spikes in $\xi_r$ and $\xi_h$ at the stellar surface should be interpreted with caution, since they can be artifacts associated with the treatment of the outermost radiative layer. The difference between the full calculation and the radiative-diffusion–only case is negligible. Comparison with the analytical equilibrium tide shows good agreement at the stellar surface, as expected, since the equilibrium tide dominates in the outer layers. {Additional comparison with the non-adiabatic calculations of \citet{Bunting2019} would be worthwhile in future work.}

To evaluate the radial wavenumber $k_r$ from standard \texttt{GYRE} outputs, we adopt the usual expression
\begin{equation}
(r k_r)^2 = \left[\frac{V}{\Gamma_1} - \frac{\ell(\ell+1)}{c_1 \sigma_{m,k}^2}\right]
\left(c_1 \sigma_{m,k}^2 - A^{\star}\right),
\end{equation}
where $V$, $A^{\star}$, and $c_1$ are structure coefficients defined in \citet{Unno1989}, and $\Gamma_1$ is the adiabatic exponent. Together with the small values of $k_r\xi_r$ and $k_h\xi_h$ at the radiative–convective boundaries, these results confirm that the tidal interaction in this model remains in the linear regime. Here $k_h = \sqrt{\ell(\ell+1)}/r$ is the horizontal wavenumber.

\begin{figure}[t]
    \centering
    \includegraphics[width=1\linewidth]{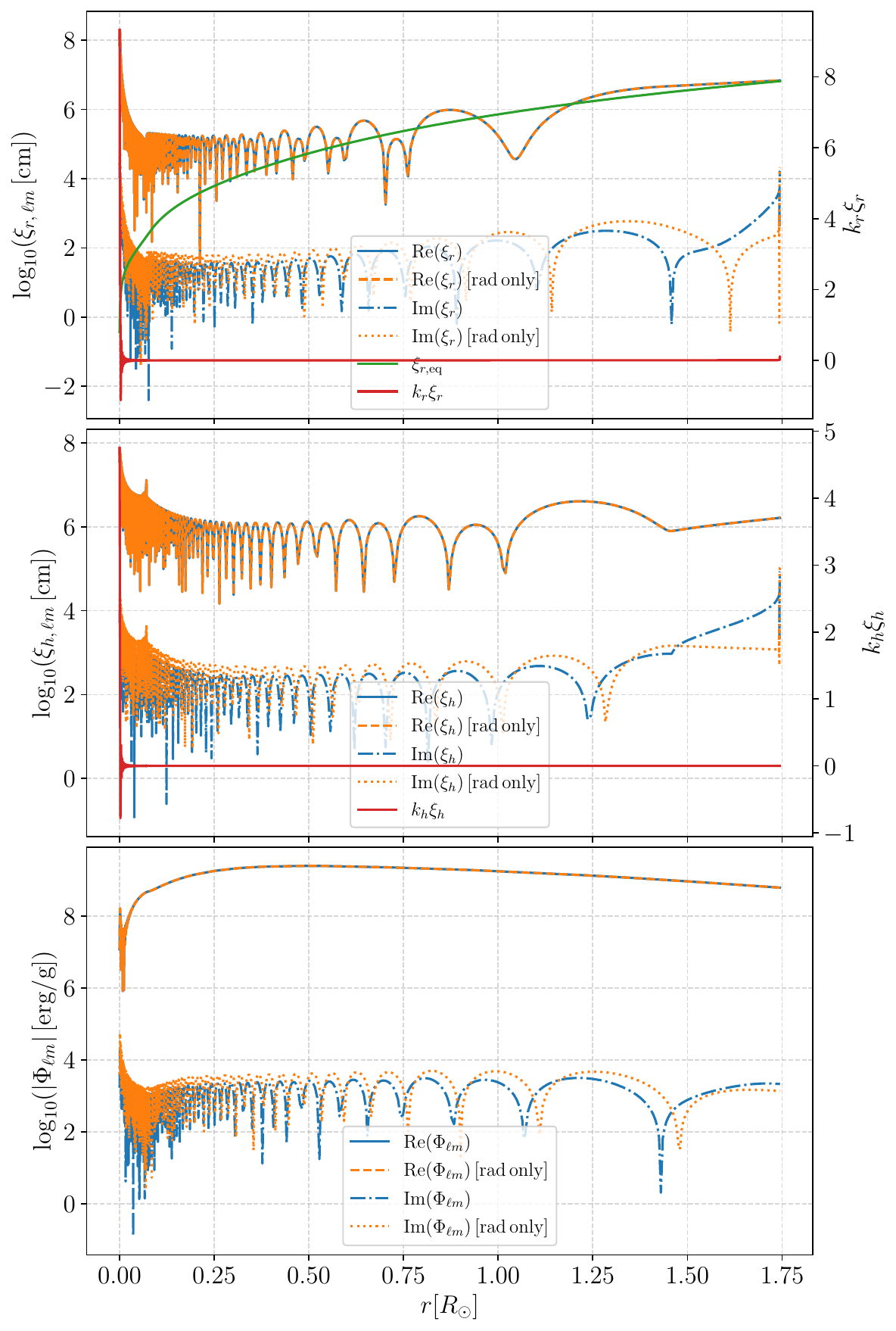}
    \caption{Same as Figure~\ref{fig:mode_analysis_k_eq_2_ell_eq_2_MS}, but for a $1.2\,M_{\odot}$ subgiant stellar model whose core is dominated by a radiative zone.}
    \label{fig:mode_analysis_k_eq_2_ell_eq_2_SG}
\end{figure}

To examine the behavior in a more evolved stellar model, we construct a subgiant using a slightly lower initial mass than our fiducial main-sequence model, with different initial metallicity and mixing-length parameter. We evolve this model until the central hydrogen abundance drops to $X_c = 10^{-5}$, at which point the stellar core becomes fully radiative. For this subgiant model shown in Figure~\ref{fig:mode_analysis_k_eq_2_ell_eq_2_SG}, the final parameters are: solar metallicity, mixing-length parameter $\alpha_{\rm MLT}=2.3$, effective temperature $T_{\rm eff} \simeq 6300\,{\rm K}$, and mean density $\bar\rho \simeq 0.32\,{\rm g\,cm^{-3}}$, which is substantially lower than in the main-sequence case. The value reflect the structural differences of an evolved star with a radiative core, which alter the mode propagation cavity and, in turn, the relative importance of nonlinear damping in the tidal response.

For the $1.2\,M_\odot$ subgiant model (Figure~\ref{fig:mode_analysis_k_eq_2_ell_eq_2_SG}), the forcing frequency does not coincide with any stellar eigenfrequency, so the system is out of resonance. As in the main-sequence case, the imaginary parts of the displacements remain much smaller than the real parts, indicating that the tidal response is close to adiabatic. The most notable feature is that the nonlinearity indicator $k_r\xi_r$ reaches values as high as $\sim8$ in the stellar core. For this 1.2 $M_{\odot}$ subgiant case, at the inner turning point $r_{\rm RCB}\simeq 0.001\,R_{\odot}$, we find $k_r\xi_r\simeq 7.81$, $\rho \simeq 8.57\times 10^2\,{\rm g\,cm^{-3}}$, and $N\simeq 6.3\times 10^{-5}\,{\rm rad\,s^{-1}}$. Substituting these values into Equation (3) of \citet{Weinberg2017}, following the moderate damping mechanism of \citet{Goodman1998}, yields a wave luminosity of order $L_{\rm w}\simeq 1.1 \times 10^{30},{\rm erg,s^{-1}}$, in excellent agreement with their result. This corresponds to orbital decay timescales of order Myr, demonstrating that subgiant structures can support much stronger tidal dissipation than their main-sequence counterparts.

It should be emphasized that our fiducial subgiant model does not fully capture the observed properties of the WASP-12 host, as already pointed out by \citet{Bailey2019}. For example, observations indicate that the star is more metal-rich than the Sun, whereas our models assume solar metallicity. In addition, a significantly enhanced mixing-length parameter is required to obtain a model that marginally agrees with the observed stellar radius and effective temperature. The primary goal of this work, however, is not to construct a perfect evolutionary model, but rather to demonstrate the consistency of our open-source non-adiabatic tidal calculations with previously well-studied systems. Importantly, our results highlight that when using \texttt{GYRE-tides} to compute the secular orbital evolution, one can also apply this approach to solar-type stars, where nonlinear tides may contribute in addition to radiative and convective dissipation, to quantify the extra tidal dissipation rate in a self-consistent way. This study further demonstrates that for stars with convective cores and radiative envelopes, our tool provides a direct way of evaluating tidal dissipation and secular evolution {in linear, non-rotating, non-adiabatic hydrodynamical models}.

\section{Conclusions\label{sec:conclusions}}
In this Letter we have revisited the tidal evolution of the WASP-12 system using direct numerical calculations with \texttt{GYRE-tides}. Our analysis was based on detailed stellar models constructed with \texttt{MESA}, and systematically explored the effects of radiative damping, convective damping, and potential nonlinear tidal responses on the orbital and spin evolution of the system. The main conclusions are as follows:

\begin{itemize}

\item \textbf{Radiative damping alone is insufficient.} When only radiative diffusion damping of dynamical tides is included, the predicted orbital decay timescales are $\sim 10^{12}$–$10^{13}$ yr, far exceeding the observed inspiral rate of WASP-12b. Resonant dips associated with high-order $g$-modes are present, but they occur only over very narrow frequency ranges; {a dedicated resonance locking analysis through numerical forced oscillation solutions is needed before this explanation can be fully excluded.}

\item \textbf{Convective damping reduces the timescales but not enough.} Incorporating tidal dissipation via turbulent viscosity decreases the secular evolution timescales by several orders of magnitude, yielding orbital decay timescales of order $10^{9}$–$10^{10}$ yr, {assuming no resonance locking occurs}. This confirms that convective damping enhances overall dissipation, yet the timescales remain markedly longer than the observed $\sim 2$ Myr decay time.

\item \textbf{Implications for additional dissipation.} {Within our non-magnetic, linear framework, the physically motivated mechanisms considered here cannot by themselves explain the observed orbital decay, implying that an additional channel is required. This extra channel could be nonlinear wave breaking. Our results remain consistent with previous indications that subgiant structures can host much stronger nonlinear dissipation than main sequence stars.}

\item \textbf{Validation of \texttt{GYRE-tides}.} The calculations confirm that \texttt{GYRE-tides}, which solves the fully non-adiabatic forced oscillation equations, reproduces the expected linear damping channels and provides a reproducible, open-source framework for analyzing tidal interactions in exoplanet systems. In addition, the code can be used to assess the onset of nonlinearity, and combined with the properties at the inner radiative boundary—offers a pathway to estimate tidal dissipation in the travelling-wave approximation caused by nonlinear damping.

\end{itemize}

Overall, the WASP-12 system continues to present a stringent test of tidal dissipation theory. {Our results suggest that a nonlinear damping mechanism may provide the dissipation necessary to explain the observed rapid orbital decay; see \citet{Duguid2024} for a linear pathway via magnetic wave conversion in strongly magnetized interiors.} Future applications to other F- to A-type stellar systems exhibiting potential decay signals {may} help clarify the role of stellar interiors in shaping the orbital evolution of hot Jupiters. Looking ahead, the Roman Galactic Bulge Time Domain Survey is expected to detect on the order of $10^5$ transiting planets, many of which will have short orbital periods and thus serve as prime candidates for tidal decay studies \citep{Carden2025}. By coupling these detections with numerical tidal response calculations from \texttt{GYRE-tides}, we can directly connect observed decay rates to the microphysics of stellar interiors. This synergy will not only sharpen constraints on the rate of planetary engulfment in the Galaxy, but also establish a lasting framework for probing tidal dissipation across diverse stellar populations.

\section*{Acknowledgments}

We thank the anonymous referee for the thoughtful comments and constructive suggestions, which helped improve the clarity and quality of this paper. This work is supported by the National Natural Science Foundation of China through the Fundamental Science Center for Nearby Galaxies (grant No. 12588202), a project based on LAMOST and FAST. M.S. and H.X. acknowledge support from the GBMF8477 grant (PI: Vicky Kalogera). R.H.D.T. acknowledges support from NASA grants 80NSSC24K0895 and 80NSSC23K1517. This work was initiated following stimulating discussions with Norm Murray during the 2025 summer workshop at the Aspen Center for Physics, where part of the research was conducted. The Center is supported by the National Science Foundation (PHY-2210452), the Simons Foundation (1161654, Troyer), and the Alfred P. Sloan Foundation (G-2024-22395).

M.S. is deeply grateful to Phil Arras for his guidance throughout this project. His early feedback was invaluable in helping M.S. identify and correct several subtleties in the calculations, and his insights greatly shaped the direction of this work. This study would not have been possible without his generous advice and encouragement. We are grateful to Zhao Guo, Chenliang Huang, Dong Lai, Daniel Lecoanet, Yoram Lithwick, Linhao Ma, Gordon Ogilvie, Fred Rasio, Xing Wei, Nevin Weinberg, Yanqin Wu and Wei Zhu for many stimulating and insightful discussions.

\software{\texttt{numpy} \citep{harris2020array}, \texttt{Matplotlib}, \texttt{pandas} \citep{reback2020pandas}, \citep{Hunter2007}, \texttt{GYRE} \citep{Townsend2013,Townsend2018,Goldstein2020,Sun2023}, \texttt{MESA} 
\citep{Paxton2011,Paxton2013,Paxton2015,Paxton2018,Paxton2019,Jermyn2023}, \texttt{MESAsdk} 
\citep{Townsend2020}}

\bibliography{wasp12.bib}{}

@ARTICLE{Paxton2011,
	author = {{Paxton}, B. and {Bildsten}, L. and {Dotter}, A. and {Herwig}, F. and 
	{Lesaffre}, P. and {Timmes}, F.},
	title = "{Modules for Experiments in Stellar Astrophysics (MESA)}",
	journal = {\apjs},
	year = 2011,
	month = jan,
	volume = 192,
	eid = {3},
	pages = {3},
	doi = {10.1088/0067-0049/192/1/3},
	adsurl = {http://adsabs.harvard.edu/abs/2011ApJS..192....3P}
}

@ARTICLE{Paxton2013,
	author = {{Paxton}, B. and {Cantiello}, M. and {Arras}, P. and {Bildsten}, L. and 
	{Brown}, E.~F. and {Dotter}, A. and {Mankovich}, C. and {Montgomery}, M.~H. and 
	{Stello}, D. and {Timmes}, F.~X. and {Townsend}, R.},
	title = "{Modules for Experiments in Stellar Astrophysics (MESA): Planets, Oscillations, Rotation, and Massive Stars}",
	journal = {\apjs},
	archivePrefix = "arXiv",
	eprint = {1301.0319},
	primaryClass = "astro-ph.SR",
	year = 2013,
	month = sep,
	volume = 208,
	eid = {4},
	pages = {4},
	doi = {10.1088/0067-0049/208/1/4},
	adsurl = {http://adsabs.harvard.edu/abs/2013ApJS..208....4P}
}

@ARTICLE{Paxton2015,
	author = {{Paxton}, B. and {Marchant}, P. and {Schwab}, J. and {Bauer}, E.~B. and 
	{Bildsten}, L. and {Cantiello}, M. and {Dessart}, L. and {Farmer}, R. and 
	{Hu}, H. and {Langer}, N. and {Townsend}, R.~H.~D. and {Townsley}, D.~M. and 
	{Timmes}, F.~X.},
	title = "{Modules for Experiments in Stellar Astrophysics (MESA): Binaries, Pulsations, and Explosions}",
	journal = {\apjs},
	year = 2015,
	month = sep,
	volume = 220,
	eid = {15},
	pages = {15},
	doi = {10.1088/0067-0049/220/1/15},
	adsurl = {http://adsabs.harvard.edu/abs/2015ApJS..220...15P}
}

@ARTICLE{Paxton2018,
       author = {{Paxton}, Bill and {Schwab}, Josiah and {Bauer}, Evan B. and
         {Bildsten}, Lars and {Blinnikov}, Sergei and {Duffell}, Paul and
         {Farmer}, R. and {Goldberg}, Jared A. and {Marchant}, Pablo and
         {Sorokina}, Elena and {Thoul}, Anne and {Townsend}, Richard H.~D. and
         {Timmes}, F.~X.},
        title = "{Modules for Experiments in Stellar Astrophysics (MESA): Convective Boundaries, Element Diffusion, and Massive Star Explosions}",
      journal = {\apjs},
         year = "2018",
        month = "Feb",
       volume = {234},
       number = {2},
          eid = {34},
        pages = {34},
          doi = {10.3847/1538-4365/aaa5a8},
archivePrefix = {arXiv},
       eprint = {1710.08424},
 primaryClass = {astro-ph.SR},
       adsurl = {https://ui.adsabs.harvard.edu/abs/2018ApJS..234...34P}
}

@ARTICLE{Paxton2019,
       author = {{Paxton}, Bill and {Smolec}, R. and {Schwab}, Josiah and {Gautschy}, A. and
         {Bildsten}, Lars and {Cantiello}, Matteo and {Dotter}, Aaron and
         {Farmer}, R. and {Goldberg}, Jared A. and {Jermyn}, Adam S. and
         {Kanbur}, S.~M. and {Marchant}, Pablo and {Thoul}, Anne and
         {Townsend}, Richard H.~D. and {Wolf}, William M. and {Zhang}, Michael and
         {Timmes}, F.~X.},
        title = "{Modules for Experiments in Stellar Astrophysics (MESA): Pulsating Variable Stars, Rotation, Convective Boundaries, and Energy Conservation}",
      journal = {\apjs},
         year = "2019",
        month = "Jul",
       volume = {243},
       number = {1},
          eid = {10},
        pages = {10},
          doi = {10.3847/1538-4365/ab2241},
       adsurl = {https://ui.adsabs.harvard.edu/abs/2019ApJS..243...10P}
}

@ARTICLE{Jermyn2023,
       author = {{Jermyn}, Adam S. and {Bauer}, Evan B. and {Schwab}, Josiah and {Farmer}, R. and {Ball}, Warrick H. and {Bellinger}, Earl P. and {Dotter}, Aaron and {Joyce}, Meridith and {Marchant}, Pablo and {Mombarg}, Joey S.~G. and {Wolf}, William M. and {Sunny Wong}, Tin Long and {Cinquegrana}, Giulia C. and {Farrell}, Eoin and {Smolec}, R. and {Thoul}, Anne and {Cantiello}, Matteo and {Herwig}, Falk and {Toloza}, Odette and {Bildsten}, Lars and {Townsend}, Richard H.~D. and {Timmes}, F.~X.},
        title = "{Modules for Experiments in Stellar Astrophysics (MESA): Time-dependent Convection, Energy Conservation, Automatic Differentiation, and Infrastructure}",
      journal = {\apjs},
         year = 2023,
        month = mar,
       volume = {265},
       number = {1},
          eid = {15},
        pages = {15},
          doi = {10.3847/1538-4365/acae8d},
archivePrefix = {arXiv},
       eprint = {2208.03651},
 primaryClass = {astro-ph.SR},
       adsurl = {https://ui.adsabs.harvard.edu/abs/2023ApJS..265...15J}
}

@ARTICLE{Townsend2013,
       author = {{Townsend}, R.~H.~D. and {Teitler}, S.~A.},
        title = "{GYRE: an open-source stellar oscillation code based on a new Magnus Multiple Shooting scheme}",
      journal = {\mnras},
         year = 2013,
        month = nov,
       volume = {435},
       number = {4},
        pages = {3406-3418},
          doi = {10.1093/mnras/stt1533},
archivePrefix = {arXiv},
       eprint = {1308.2965},
 primaryClass = {astro-ph.SR},
       adsurl = {https://ui.adsabs.harvard.edu/abs/2013MNRAS.435.3406T}
}

@ARTICLE{Townsend2018,
       author = {{Townsend}, R.~H.~D. and {Goldstein}, J. and {Zweibel}, E.~G.},
        title = "{Angular momentum transport by heat-driven g-modes in slowly pulsating B stars}",
      journal = {\mnras},
         year = 2018,
        month = mar,
       volume = {475},
       number = {1},
        pages = {879-893},
          doi = {10.1093/mnras/stx3142},
archivePrefix = {arXiv},
       eprint = {1712.02420},
 primaryClass = {astro-ph.SR},
       adsurl = {https://ui.adsabs.harvard.edu/abs/2018MNRAS.475..879T}
}

@ARTICLE{Goldstein2020,
       author = {{Goldstein}, J. and {Townsend}, R.~H.~D.},
        title = "{The Contour Method: a New Approach to Finding Modes of Nonadiabatic Stellar Pulsations}",
      journal = {\apj},
         year = 2020,
        month = aug,
       volume = {899},
       number = {2},
          eid = {116},
        pages = {116},
          doi = {10.3847/1538-4357/aba748},
archivePrefix = {arXiv},
       eprint = {2006.13223},
 primaryClass = {astro-ph.SR},
       adsurl = {https://ui.adsabs.harvard.edu/abs/2020ApJ...899..116G}
}

@ARTICLE{Sun2023,
       author = {{Sun}, Meng and {Townsend}, R.~H.~D. and {Guo}, Zhao},
        title = "{gyre\_tides: Modeling Binary Tides within the GYRE Stellar Oscillation Code}",
      journal = {\apj},
         year = 2023,
        month = mar,
       volume = {945},
       number = {1},
          eid = {43},
        pages = {43},
          doi = {10.3847/1538-4357/acb33a},
archivePrefix = {arXiv},
       eprint = {2301.06599},
 primaryClass = {astro-ph.SR},
       adsurl = {https://ui.adsabs.harvard.edu/abs/2023ApJ...945...43S}
}

@Article{harris2020array,
 title         = {Array programming with {NumPy}},
 author        = {Charles R. Harris and K. Jarrod Millman and St{\'{e}}fan J.
                 van der Walt and Ralf Gommers and Pauli Virtanen and David
                 Cournapeau and Eric Wieser and Julian Taylor and Sebastian
                 Berg and Nathaniel J. Smith and Robert Kern and Matti Picus
                 and Stephan Hoyer and Marten H. van Kerkwijk and Matthew
                 Brett and Allan Haldane and Jaime Fern{\'{a}}ndez del
                 R{\'{i}}o and Mark Wiebe and Pearu Peterson and Pierre
                 G{\'{e}}rard-Marchant and Kevin Sheppard and Tyler Reddy and
                 Warren Weckesser and Hameer Abbasi and Christoph Gohlke and
                 Travis E. Oliphant},
 year          = {2020},
 month         = sep,
 journal       = {Nature},
 volume        = {585},
 number        = {7825},
 pages         = {357--362},
 doi           = {10.1038/s41586-020-2649-2},
 publisher     = {Springer Science and Business Media {LLC}},
 url           = {https://doi.org/10.1038/s41586-020-2649-2}
}

@Article{Hunter2007,
  Author    = {Hunter, J. D.},
  Title     = {Matplotlib: A 2D graphics environment},
  Journal   = {Computing in Science \& Engineering},
  Volume    = {9},
  Number    = {3},
  Pages     = {90--95},
  publisher = {IEEE COMPUTER SOC},
  doi       = {10.1109/MCSE.2007.55},
  year      = 2007
}

@ARTICLE{Weinberg2017,
       author = {{Weinberg}, Nevin N. and {Sun}, Meng and {Arras}, Phil and {Essick}, Reed},
        title = "{Tidal Dissipation in WASP-12}",
      journal = {\apjl},
         year = 2017,
        month = nov,
       volume = {849},
       number = {1},
          eid = {L11},
        pages = {L11},
          doi = {10.3847/2041-8213/aa9113},
archivePrefix = {arXiv},
       eprint = {1710.00858},
 primaryClass = {astro-ph.EP},
       adsurl = {https://ui.adsabs.harvard.edu/abs/2017ApJ...849L..11W}
}

@ARTICLE{Millholland2018,
       author = {{Millholland}, Sarah and {Laughlin}, Gregory},
        title = "{Obliquity Tides May Drive WASP-12b{\textquoteright}s Rapid Orbital Decay}",
      journal = {\apjl},
         year = 2018,
        month = dec,
       volume = {869},
       number = {1},
          eid = {L15},
        pages = {L15},
          doi = {10.3847/2041-8213/aaedb1},
archivePrefix = {arXiv},
       eprint = {1812.01624},
 primaryClass = {astro-ph.EP},
       adsurl = {https://ui.adsabs.harvard.edu/abs/2018ApJ...869L..15M}
}

@ARTICLE{Bailey2019,
       author = {{Bailey}, Avery and {Goodman}, Jeremy},
        title = "{Understanding WASP-12b}",
      journal = {\mnras},
         year = 2019,
        month = jan,
       volume = {482},
       number = {2},
        pages = {1872-1882},
          doi = {10.1093/mnras/sty2805},
archivePrefix = {arXiv},
       eprint = {1808.00052},
 primaryClass = {astro-ph.EP},
       adsurl = {https://ui.adsabs.harvard.edu/abs/2019MNRAS.482.1872B}
}

@ARTICLE{Willems2010,
       author = {{Willems}, B. and {Deloye}, C.~J. and {Kalogera}, V.},
        title = "{Energy Dissipation Through Quasi-static Tides in White Dwarf Binaries}",
      journal = {\apj},
         year = 2010,
        month = apr,
       volume = {713},
       number = {1},
        pages = {239-256},
          doi = {10.1088/0004-637X/713/1/239},
archivePrefix = {arXiv},
       eprint = {0904.1953},
 primaryClass = {astro-ph.SR},
       adsurl = {https://ui.adsabs.harvard.edu/abs/2010ApJ...713..239W}
}

@ARTICLE{Sun2018,
       author = {{Sun}, M. and {Arras}, P. and {Weinberg}, N.~N. and {Troup}, N.~W. and {Majewski}, S.~R.},
        title = "{Orbital decay in binaries containing post-main-sequence stars}",
      journal = {\mnras},
         year = 2018,
        month = dec,
       volume = {481},
       number = {3},
        pages = {4077-4092},
          doi = {10.1093/mnras/sty2464},
archivePrefix = {arXiv},
       eprint = {1809.02187},
 primaryClass = {astro-ph.SR},
       adsurl = {https://ui.adsabs.harvard.edu/abs/2018MNRAS.481.4077S}
}

@ARTICLE{Leonardi2024,
       author = {{Leonardi}, P. and {Nascimbeni}, V. and {Granata}, V. and {Malavolta}, L. and {Borsato}, L. and {Biazzo}, K. and {Lanza}, A.~F. and {Desidera}, S. and {Piotto}, G. and {Nardiello}, D. and {Damasso}, M. and {Cunial}, A. and {Bedin}, L.~R.},
        title = "{TASTE. V. A new ground-based investigation of orbital decay in the ultra-hot Jupiter WASP-12b}",
      journal = {\aap},
         year = 2024,
        month = jun,
       volume = {686},
          eid = {A84},
        pages = {A84},
          doi = {10.1051/0004-6361/202348363},
archivePrefix = {arXiv},
       eprint = {2402.12120},
 primaryClass = {astro-ph.EP},
       adsurl = {https://ui.adsabs.harvard.edu/abs/2024A&A...686A..84L}
}

@ARTICLE{Stassun2019,
       author = {{Stassun}, Keivan G. and {Oelkers}, Ryan J. and {Paegert}, Martin and {Torres}, Guillermo and {Pepper}, Joshua and {De Lee}, Nathan and {Collins}, Kevin and {Latham}, David W. and {Muirhead}, Philip S. and {Chittidi}, Jay and {Rojas-Ayala}, B{\'a}rbara and {Fleming}, Scott W. and {Rose}, Mark E. and {Tenenbaum}, Peter and {Ting}, Eric B. and {Kane}, Stephen R. and {Barclay}, Thomas and {Bean}, Jacob L. and {Brassuer}, C.~E. and {Charbonneau}, David and {Ge}, Jian and {Lissauer}, Jack J. and {Mann}, Andrew W. and {McLean}, Brian and {Mullally}, Susan and {Narita}, Norio and {Plavchan}, Peter and {Ricker}, George R. and {Sasselov}, Dimitar and {Seager}, S. and {Sharma}, Sanjib and {Shiao}, Bernie and {Sozzetti}, Alessandro and {Stello}, Dennis and {Vanderspek}, Roland and {Wallace}, Geoff and {Winn}, Joshua N.},
        title = "{The Revised TESS Input Catalog and Candidate Target List}",
      journal = {\aj},
         year = 2019,
        month = oct,
       volume = {158},
       number = {4},
          eid = {138},
        pages = {138},
          doi = {10.3847/1538-3881/ab3467},
archivePrefix = {arXiv},
       eprint = {1905.10694},
 primaryClass = {astro-ph.SR},
       adsurl = {https://ui.adsabs.harvard.edu/abs/2019AJ....158..138S}
}

@ARTICLE{Chakrabarty2019,
       author = {{Chakrabarty}, Aritra and {Sengupta}, Sujan},
        title = "{Precise Photometric Transit Follow-up Observations of Five Close-in Exoplanets: Update on Their Physical Properties}",
      journal = {\aj},
         year = 2019,
        month = jul,
       volume = {158},
       number = {1},
          eid = {39},
        pages = {39},
          doi = {10.3847/1538-3881/ab24dd},
archivePrefix = {arXiv},
       eprint = {1905.11258},
 primaryClass = {astro-ph.EP},
       adsurl = {https://ui.adsabs.harvard.edu/abs/2019AJ....158...39C}
}

@ARTICLE{Ozturk2019,
       author = {{{\"O}zt{\"u}rk}, O{\v{g}}uz and {Erdem}, Ahmet},
        title = "{New photometric analysis of five exoplanets: CoRoT-2b, HAT-P-12b, TrES-2b, WASP-12b, and WASP-52b}",
      journal = {\mnras},
         year = 2019,
        month = jun,
       volume = {486},
       number = {2},
        pages = {2290-2307},
          doi = {10.1093/mnras/stz747},
       adsurl = {https://ui.adsabs.harvard.edu/abs/2019MNRAS.486.2290O}
}

@ARTICLE{Gaia2018,
       author = {{Gaia Collaboration} and {Brown}, A.~G.~A. and {Vallenari}, A. and {Prusti}, T. and {de Bruijne}, J.~H.~J. and {Babusiaux}, C. and {Bailer-Jones}, C.~A.~L. and {Biermann}, M. and {Evans}, D.~W. and {Eyer}, L. and {Jansen}, F. and {Jordi}, C. and {Klioner}, S.~A. and {Lammers}, U. and {Lindegren}, L. and {Luri}, X. and {Mignard}, F. and {Panem}, C. and {Pourbaix}, D. and {Randich}, S. and {Sartoretti}, P. and {Siddiqui}, H.~I. and {Soubiran}, C. and {van Leeuwen}, F. and {Walton}, N.~A. and {Arenou}, F. and {Bastian}, U. and {Cropper}, M. and {Drimmel}, R. and {Katz}, D. and {Lattanzi}, M.~G. and {Bakker}, J. and {Cacciari}, C. and {Casta{\~n}eda}, J. and {Chaoul}, L. and {Cheek}, N. and {De Angeli}, F. and {Fabricius}, C. and {Guerra}, R. and {Holl}, B. and {Masana}, E. and {Messineo}, R. and {Mowlavi}, N. and {Nienartowicz}, K. and {Panuzzo}, P. and {Portell}, J. and {Riello}, M. and {Seabroke}, G.~M. and {Tanga}, P. and {Th{\'e}venin}, F. and {Gracia-Abril}, G. and {Comoretto}, G. and {Garcia-Reinaldos}, M. and {Teyssier}, D. and {Altmann}, M. and {Andrae}, R. and {Audard}, M. and {Bellas-Velidis}, I. and {Benson}, K. and {Berthier}, J. and {Blomme}, R. and {Burgess}, P. and {Busso}, G. and {Carry}, B. and {Cellino}, A. and {Clementini}, G. and {Clotet}, M. and {Creevey}, O. and {Davidson}, M. and {De Ridder}, J. and {Delchambre}, L. and {Dell'Oro}, A. and {Ducourant}, C. and {Fern{\'a}ndez-Hern{\'a}ndez}, J. and {Fouesneau}, M. and {Fr{\'e}mat}, Y. and {Galluccio}, L. and {Garc{\'\i}a-Torres}, M. and {Gonz{\'a}lez-N{\'u}{\~n}ez}, J. and {Gonz{\'a}lez-Vidal}, J.~J. and {Gosset}, E. and {Guy}, L.~P. and {Halbwachs}, J. -L. and {Hambly}, N.~C. and {Harrison}, D.~L. and {Hern{\'a}ndez}, J. and {Hestroffer}, D. and {Hodgkin}, S.~T. and {Hutton}, A. and {Jasniewicz}, G. and {Jean-Antoine-Piccolo}, A. and {Jordan}, S. and {Korn}, A.~J. and {Krone-Martins}, A. and {Lanzafame}, A.~C. and {Lebzelter}, T. and {L{\"o}ffler}, W. and {Manteiga}, M. and {Marrese}, P.~M. and {Mart{\'\i}n-Fleitas}, J.~M. and {Moitinho}, A. and {Mora}, A. and {Muinonen}, K. and {Osinde}, J. and {Pancino}, E. and {Pauwels}, T. and {Petit}, J. -M. and {Recio-Blanco}, A. and {Richards}, P.~J. and {Rimoldini}, L. and {Robin}, A.~C. and {Sarro}, L.~M. and {Siopis}, C. and {Smith}, M. and {Sozzetti}, A. and {S{\"u}veges}, M. and {Torra}, J. and {van Reeven}, W. and {Abbas}, U. and {Abreu Aramburu}, A. and {Accart}, S. and {Aerts}, C. and {Altavilla}, G. and {{\'A}lvarez}, M.~A. and {Alvarez}, R. and {Alves}, J. and {Anderson}, R.~I. and {Andrei}, A.~H. and {Anglada Varela}, E. and {Antiche}, E. and {Antoja}, T. and {Arcay}, B. and {Astraatmadja}, T.~L. and {Bach}, N. and {Baker}, S.~G. and {Balaguer-N{\'u}{\~n}ez}, L. and {Balm}, P. and {Barache}, C. and {Barata}, C. and {Barbato}, D. and {Barblan}, F. and {Barklem}, P.~S. and {Barrado}, D. and {Barros}, M. and {Barstow}, M.~A. and {Bartholom{\'e} Mu{\~n}oz}, S. and {Bassilana}, J. -L. and {Becciani}, U. and {Bellazzini}, M. and {Berihuete}, A. and {Bertone}, S. and {Bianchi}, L. and {Bienaym{\'e}}, O. and {Blanco-Cuaresma}, S. and {Boch}, T. and {Boeche}, C. and {Bombrun}, A. and {Borrachero}, R. and {Bossini}, D. and {Bouquillon}, S. and {Bourda}, G. and {Bragaglia}, A. and {Bramante}, L. and {Breddels}, M.~A. and {Bressan}, A. and {Brouillet}, N. and {Br{\"u}semeister}, T. and {Brugaletta}, E. and {Bucciarelli}, B. and {Burlacu}, A. and {Busonero}, D. and {Butkevich}, A.~G. and {Buzzi}, R. and {Caffau}, E. and {Cancelliere}, R. and {Cannizzaro}, G. and {Cantat-Gaudin}, T. and {Carballo}, R. and {Carlucci}, T. and {Carrasco}, J.~M. and {Casamiquela}, L. and {Castellani}, M. and {Castro-Ginard}, A. and {Charlot}, P. and {Chemin}, L. and {Chiavassa}, A. and {Cocozza}, G. and {Costigan}, G. and {Cowell}, S. and {Crifo}, F. and {Crosta}, M. and {Crowley}, C. and {Cuypers}, J. and {Dafonte}, C. and {Damerdji}, Y. and {Dapergolas}, A. and {David}, P. and {David}, M. and {de Laverny}, P. and {De Luise}, F.},
        title = "{Gaia Data Release 2. Summary of the contents and survey properties}",
      journal = {\aap},
         year = 2018,
        month = aug,
       volume = {616},
          eid = {A1},
        pages = {A1},
          doi = {10.1051/0004-6361/201833051},
archivePrefix = {arXiv},
       eprint = {1804.09365},
 primaryClass = {astro-ph.GA},
       adsurl = {https://ui.adsabs.harvard.edu/abs/2018A&A...616A...1G}
}

@ARTICLE{Bonomo2017,
       author = {{Bonomo}, A.~S. and {Desidera}, S. and {Benatti}, S. and {Borsa}, F. and {Crespi}, S. and {Damasso}, M. and {Lanza}, A.~F. and {Sozzetti}, A. and {Lodato}, G. and {Marzari}, F. and {Boccato}, C. and {Claudi}, R.~U. and {Cosentino}, R. and {Covino}, E. and {Gratton}, R. and {Maggio}, A. and {Micela}, G. and {Molinari}, E. and {Pagano}, I. and {Piotto}, G. and {Poretti}, E. and {Smareglia}, R. and {Affer}, L. and {Biazzo}, K. and {Bignamini}, A. and {Esposito}, M. and {Giacobbe}, P. and {H{\'e}brard}, G. and {Malavolta}, L. and {Maldonado}, J. and {Mancini}, L. and {Martinez Fiorenzano}, A. and {Masiero}, S. and {Nascimbeni}, V. and {Pedani}, M. and {Rainer}, M. and {Scandariato}, G.},
        title = "{The GAPS Programme with HARPS-N at TNG . XIV. Investigating giant planet migration history via improved eccentricity and mass determination for 231 transiting planets}",
      journal = {\aap},
         year = 2017,
        month = jun,
       volume = {602},
          eid = {A107},
        pages = {A107},
          doi = {10.1051/0004-6361/201629882},
archivePrefix = {arXiv},
       eprint = {1704.00373},
 primaryClass = {astro-ph.EP},
       adsurl = {https://ui.adsabs.harvard.edu/abs/2017A&A...602A.107B}
}

@ARTICLE{Stassun2017,
       author = {{Stassun}, Keivan G. and {Collins}, Karen A. and {Gaudi}, B. Scott},
        title = "{Accurate Empirical Radii and Masses of Planets and Their Host Stars with Gaia Parallaxes}",
      journal = {\aj},
         year = 2017,
        month = mar,
       volume = {153},
       number = {3},
          eid = {136},
        pages = {136},
          doi = {10.3847/1538-3881/aa5df3},
archivePrefix = {arXiv},
       eprint = {1609.04389},
 primaryClass = {astro-ph.EP},
       adsurl = {https://ui.adsabs.harvard.edu/abs/2017AJ....153..136S}
}

@ARTICLE{Collins2017,
       author = {{Collins}, Karen A. and {Kielkopf}, John F. and {Stassun}, Keivan G.},
        title = "{Transit Timing Variation Measurements of WASP-12b and Qatar-1b: No Evidence Of Additional Planets}",
      journal = {\aj},
         year = 2017,
        month = feb,
       volume = {153},
       number = {2},
          eid = {78},
        pages = {78},
          doi = {10.3847/1538-3881/153/2/78},
archivePrefix = {arXiv},
       eprint = {1512.00464},
 primaryClass = {astro-ph.EP},
       adsurl = {https://ui.adsabs.harvard.edu/abs/2017AJ....153...78C}
}

@ARTICLE{Turner2016,
       author = {{Turner}, Jake D. and {Pearson}, Kyle A. and {Biddle}, Lauren I. and {Smart}, Brianna M. and {Zellem}, Robert T. and {Teske}, Johanna K. and {Hardegree-Ullman}, Kevin K. and {Griffith}, Caitlin C. and {Leiter}, Robin M. and {Cates}, Ian T. and {Nieberding}, Megan N. and {Smith}, Carter-Thaxton W. and {Thompson}, Robert M. and {Hofmann}, Ryan and {Berube}, Michael P. and {Nguyen}, Chi H. and {Small}, Lindsay C. and {Guvenen}, Blythe C. and {Richardson}, Logan and {McGraw}, Allison and {Raphael}, Brandon and {Crawford}, Benjamin E. and {Robertson}, Amy N. and {Tombleson}, Ryan and {Carleton}, Timothy M. and {Towner}, Allison P.~M. and {Walker-LaFollette}, Amanda M. and {Hume}, Jeffrey R. and {Watson}, Zachary T. and {Jones}, Christen K. and {Lichtenberger}, Matthew J. and {Hoglund}, Shelby R. and {Cook}, Kendall L. and {Crossen}, Cory A. and {Jorgensen}, Curtis R. and {Romine}, James M. and {Thompson}, Alejandro R. and {Villegas}, Christian F. and {Wilson}, Ashley A. and {Sanford}, Brent and {Taylor}, Joanna M. and {Henz}, Triana N.},
        title = "{Ground-based near-UV observations of 15 transiting exoplanets: constraints on their atmospheres and no evidence for asymmetrical transits}",
      journal = {\mnras},
         year = 2016,
        month = jun,
       volume = {459},
       number = {1},
        pages = {789-819},
          doi = {10.1093/mnras/stw574},
archivePrefix = {arXiv},
       eprint = {1603.02587},
 primaryClass = {astro-ph.EP},
       adsurl = {https://ui.adsabs.harvard.edu/abs/2016MNRAS.459..789T}
}

@ARTICLE{Knutson2014,
       author = {{Knutson}, Heather A. and {Fulton}, Benjamin J. and {Montet}, Benjamin T. and {Kao}, Melodie and {Ngo}, Henry and {Howard}, Andrew W. and {Crepp}, Justin R. and {Hinkley}, Sasha and {Bakos}, Gaspar {\'A}. and {Batygin}, Konstantin and {Johnson}, John Asher and {Morton}, Timothy D. and {Muirhead}, Philip S.},
        title = "{Friends of Hot Jupiters. I. A Radial Velocity Search for Massive, Long-period Companions to Close-in Gas Giant Planets}",
      journal = {\apj},
         year = 2014,
        month = apr,
       volume = {785},
       number = {2},
          eid = {126},
        pages = {126},
          doi = {10.1088/0004-637X/785/2/126},
archivePrefix = {arXiv},
       eprint = {1312.2954},
 primaryClass = {astro-ph.EP},
       adsurl = {https://ui.adsabs.harvard.edu/abs/2014ApJ...785..126K}
}

@ARTICLE{Mortier2013,
       author = {{Mortier}, A. and {Santos}, N.~C. and {Sousa}, S.~G. and {Fernandes}, J.~M. and {Adibekyan}, V. Zh. and {Delgado Mena}, E. and {Montalto}, M. and {Israelian}, G.},
        title = "{New and updated stellar parameters for 90 transit hosts. The effect of the surface gravity}",
      journal = {\aap},
         year = 2013,
        month = oct,
       volume = {558},
          eid = {A106},
        pages = {A106},
          doi = {10.1051/0004-6361/201322240},
archivePrefix = {arXiv},
       eprint = {1309.1998},
 primaryClass = {astro-ph.EP},
       adsurl = {https://ui.adsabs.harvard.edu/abs/2013A&A...558A.106M}
}

@ARTICLE{Southworth2012,
       author = {{Southworth}, John},
        title = "{Homogeneous studies of transiting extrasolar planets - V. New results for 38 planets}",
      journal = {\mnras},
         year = 2012,
        month = oct,
       volume = {426},
       number = {2},
        pages = {1291-1323},
          doi = {10.1111/j.1365-2966.2012.21756.x},
archivePrefix = {arXiv},
       eprint = {1207.5796},
 primaryClass = {astro-ph.EP},
       adsurl = {https://ui.adsabs.harvard.edu/abs/2012MNRAS.426.1291S}
}

@ARTICLE{Hebb2009,
       author = {{Hebb}, L. and {Collier-Cameron}, A. and {Loeillet}, B. and {Pollacco}, D. and {H{\'e}brard}, G. and {Street}, R.~A. and {Bouchy}, F. and {Stempels}, H.~C. and {Moutou}, C. and {Simpson}, E. and {Udry}, S. and {Joshi}, Y.~C. and {West}, R.~G. and {Skillen}, I. and {Wilson}, D.~M. and {McDonald}, I. and {Gibson}, N.~P. and {Aigrain}, S. and {Anderson}, D.~R. and {Benn}, C.~R. and {Christian}, D.~J. and {Enoch}, B. and {Haswell}, C.~A. and {Hellier}, C. and {Horne}, K. and {Irwin}, J. and {Lister}, T.~A. and {Maxted}, P. and {Mayor}, M. and {Norton}, A.~J. and {Parley}, N. and {Pont}, F. and {Queloz}, D. and {Smalley}, B. and {Wheatley}, P.~J.},
        title = "{WASP-12b: The Hottest Transiting Extrasolar Planet Yet Discovered}",
      journal = {\apj},
         year = 2009,
        month = mar,
       volume = {693},
       number = {2},
        pages = {1920-1928},
          doi = {10.1088/0004-637X/693/2/1920},
archivePrefix = {arXiv},
       eprint = {0812.3240},
 primaryClass = {astro-ph},
       adsurl = {https://ui.adsabs.harvard.edu/abs/2009ApJ...693.1920H}
}

@ARTICLE{Maciejewski2016,
       author = {{Maciejewski}, G. and {Dimitrov}, D. and {Fern{\'a}ndez}, M. and {Sota}, A. and {Nowak}, G. and {Ohlert}, J. and {Nikolov}, G. and {Bukowiecki}, {\L}. and {Hinse}, T.~C. and {Pall{\'e}}, E. and {Tingley}, B. and {Kjurkchieva}, D. and {Lee}, J.~W. and {Lee}, C. -U.},
        title = "{Departure from the constant-period ephemeris for the transiting exoplanet WASP-12}",
      journal = {\aap},
         year = 2016,
        month = apr,
       volume = {588},
          eid = {L6},
        pages = {L6},
          doi = {10.1051/0004-6361/201628312},
archivePrefix = {arXiv},
       eprint = {1602.09055},
 primaryClass = {astro-ph.EP},
       adsurl = {https://ui.adsabs.harvard.edu/abs/2016A&A...588L...6M}
}

@ARTICLE{Patra2017,
       author = {{Patra}, Kishore C. and {Winn}, Joshua N. and {Holman}, Matthew J. and {Yu}, Liang and {Deming}, Drake and {Dai}, Fei},
        title = "{The Apparently Decaying Orbit of WASP-12b}",
      journal = {\aj},
         year = 2017,
        month = jul,
       volume = {154},
       number = {1},
          eid = {4},
        pages = {4},
          doi = {10.3847/1538-3881/aa6d75},
archivePrefix = {arXiv},
       eprint = {1703.06582},
 primaryClass = {astro-ph.EP},
       adsurl = {https://ui.adsabs.harvard.edu/abs/2017AJ....154....4P}
}

@ARTICLE{Yee2020,
       author = {{Yee}, Samuel W. and {Winn}, Joshua N. and {Knutson}, Heather A. and {Patra}, Kishore C. and {Vissapragada}, Shreyas and {Zhang}, Michael M. and {Holman}, Matthew J. and {Shporer}, Avi and {Wright}, Jason T.},
        title = "{The Orbit of WASP-12b Is Decaying}",
      journal = {\apjl},
         year = 2020,
        month = jan,
       volume = {888},
       number = {1},
          eid = {L5},
        pages = {L5},
          doi = {10.3847/2041-8213/ab5c16},
archivePrefix = {arXiv},
       eprint = {1911.09131},
 primaryClass = {astro-ph.EP},
       adsurl = {https://ui.adsabs.harvard.edu/abs/2020ApJ...888L...5Y}
}

@ARTICLE{Duguid2024,
       author = {{Duguid}, Craig D. and {de Vries}, Nils B. and {Lecoanet}, Daniel and {Barker}, Adrian J.},
        title = "{An Efficient Tidal Dissipation Mechanism via Stellar Magnetic Fields}",
      journal = {\apjl},
         year = 2024,
        month = may,
       volume = {966},
       number = {1},
          eid = {L14},
        pages = {L14},
          doi = {10.3847/2041-8213/ad3c40},
archivePrefix = {arXiv},
       eprint = {2404.07085},
 primaryClass = {astro-ph.SR},
       adsurl = {https://ui.adsabs.harvard.edu/abs/2024ApJ...966L..14D}
}

@ARTICLE{Chernov2017,
       author = {{Chernov}, S.~V. and {Ivanov}, P.~B. and {Papaloizou}, J.~C.~B.},
        title = "{Dynamical tides in exoplanetary systems containing hot Jupiters: confronting theory and observations}",
      journal = {\mnras},
         year = 2017,
        month = sep,
       volume = {470},
       number = {2},
        pages = {2054-2068},
          doi = {10.1093/mnras/stx1234},
archivePrefix = {arXiv},
       eprint = {1705.06699},
 primaryClass = {astro-ph.SR},
       adsurl = {https://ui.adsabs.harvard.edu/abs/2017MNRAS.470.2054C}
}

@ARTICLE{Choi2016,
       author = {{Choi}, Jieun and {Dotter}, Aaron and {Conroy}, Charlie and {Cantiello}, Matteo and {Paxton}, Bill and {Johnson}, Benjamin D.},
        title = "{Mesa Isochrones and Stellar Tracks (MIST). I. Solar-scaled Models}",
      journal = {\apj},
         year = 2016,
        month = jun,
       volume = {823},
       number = {2},
          eid = {102},
        pages = {102},
          doi = {10.3847/0004-637X/823/2/102},
archivePrefix = {arXiv},
       eprint = {1604.08592},
 primaryClass = {astro-ph.SR},
       adsurl = {https://ui.adsabs.harvard.edu/abs/2016ApJ...823..102C}
}

@ARTICLE{Terquem1998,
       author = {{Terquem}, C. and {Papaloizou}, J.~C.~B. and {Nelson}, R.~P. and {Lin}, D.~N.~C.},
        title = "{On the Tidal Interaction of a Solar-Type Star with an Orbiting Companion: Excitation of g-Mode Oscillation and Orbital Evolution}",
      journal = {\apj},
         year = 1998,
        month = aug,
       volume = {502},
       number = {2},
        pages = {788-801},
          doi = {10.1086/305927},
archivePrefix = {arXiv},
       eprint = {astro-ph/9801280},
 primaryClass = {astro-ph},
       adsurl = {https://ui.adsabs.harvard.edu/abs/1998ApJ...502..788T}
}

@ARTICLE{Cowling1941,
       author = {{Cowling}, T.~G.},
        title = "{The non-radial oscillations of polytropic stars}",
      journal = {\mnras},
         year = 1941,
        month = jan,
       volume = {101},
        pages = {367},
          doi = {10.1093/mnras/101.8.367},
       adsurl = {https://ui.adsabs.harvard.edu/abs/1941MNRAS.101..367C}
}

@BOOK{Unno1989,
       author = {{Unno}, Wasaburo and {Osaki}, Yoji and {Ando}, Hiroyasu and {Saio}, H. and {Shibahashi}, H.},
        title = "{Nonradial oscillations of stars}",
         year = 1989,
       adsurl = {https://ui.adsabs.harvard.edu/abs/1989nos..book.....U}
}

@ARTICLE{Weinberg2024,
       author = {{Weinberg}, Nevin N. and {Davachi}, Niyousha and {Essick}, Reed and {Yu}, Hang and {Arras}, Phil and {Belland}, Brent},
        title = "{Orbital Decay of Hot Jupiters due to Weakly Nonlinear Tidal Dissipation}",
      journal = {\apj},
         year = 2024,
        month = jan,
       volume = {960},
       number = {1},
          eid = {50},
        pages = {50},
          doi = {10.3847/1538-4357/ad05c9},
archivePrefix = {arXiv},
       eprint = {2305.11974},
 primaryClass = {astro-ph.EP},
       adsurl = {https://ui.adsabs.harvard.edu/abs/2024ApJ...960...50W}
}

@ARTICLE{Verbunt1995,
       author = {{Verbunt}, F. and {Phinney}, E.~S.},
        title = "{Tidal circularization and the eccentricity of binaries containing giant stars.}",
      journal = {\aap},
         year = 1995,
        month = apr,
       volume = {296},
        pages = {709},
       adsurl = {https://ui.adsabs.harvard.edu/abs/1995A&A...296..709V}
}

@ARTICLE{Zahn1989,
       author = {{Zahn}, J. -P.},
        title = "{Tidal evolution of close binary stars. I - Revisiting the theory of the equilibrium tide}",
      journal = {\aap},
         year = 1989,
        month = aug,
       volume = {220},
       number = {1-2},
        pages = {112-116},
       adsurl = {https://ui.adsabs.harvard.edu/abs/1989A&A...220..112Z}
}

@ARTICLE{Pfahl2008,
       author = {{Pfahl}, Eric and {Arras}, Phil and {Paxton}, Bill},
        title = "{Ellipsoidal Oscillations Induced by Substellar Companions: A Prospect for the Kepler Mission}",
      journal = {\apj},
         year = 2008,
        month = may,
       volume = {679},
       number = {1},
        pages = {783-796},
          doi = {10.1086/586878},
archivePrefix = {arXiv},
       eprint = {0704.1910},
 primaryClass = {astro-ph},
       adsurl = {https://ui.adsabs.harvard.edu/abs/2008ApJ...679..783P}
}

@ARTICLE{Goodman1998,
       author = {{Goodman}, Jeremy and {Dickson}, Eric S.},
        title = "{Dynamical Tide in Solar-Type Binaries}",
      journal = {\apj},
         year = 1998,
        month = nov,
       volume = {507},
       number = {2},
        pages = {938-944},
          doi = {10.1086/306348},
archivePrefix = {arXiv},
       eprint = {astro-ph/9801289},
 primaryClass = {astro-ph},
       adsurl = {https://ui.adsabs.harvard.edu/abs/1998ApJ...507..938G}
}

@ARTICLE{Carden2025,
       author = {{Carden}, Kylee and {Gaudi}, B. Scott and {Wilson}, Robert F.},
        title = "{A Short History of (Orbital) Decay: Roman's Prospects for Detecting Dying Planets}",
      journal = {\aj},
         year = 2025,
        month = aug,
       volume = {170},
       number = {2},
          eid = {93},
        pages = {93},
          doi = {10.3847/1538-3881/ade3d9},
archivePrefix = {arXiv},
       eprint = {2504.15277},
 primaryClass = {astro-ph.EP},
       adsurl = {https://ui.adsabs.harvard.edu/abs/2025AJ....170...93C}
}

@ARTICLE{Barker2020,
       author = {{Barker}, A.~J.},
        title = "{Tidal dissipation in evolving low-mass and solar-type stars with predictions for planetary orbital decay}",
      journal = {\mnras},
         year = 2020,
        month = oct,
       volume = {498},
       number = {2},
        pages = {2270-2294},
          doi = {10.1093/mnras/staa2405},
archivePrefix = {arXiv},
       eprint = {2008.03262},
 primaryClass = {astro-ph.EP},
       adsurl = {https://ui.adsabs.harvard.edu/abs/2020MNRAS.498.2270B}
}

@software{reback2020pandas,
author = {The pandas development team},
title = {pandas-dev/pandas: Pandas},
month = feb,
year = 2020,
publisher = {Zenodo},
version = {latest},
doi = {10.5281/zenodo.3509134},
url = {https://doi.org/10.5281/zenodo.3509134}
}

@ARTICLE{Lai2010,
       author = {{Lai}, Dong and {Helling}, Ch. and {van den Heuvel}, E.~P.~J.},
        title = "{Mass Transfer, Transiting Stream, and Magnetopause in Close-in Exoplanetary Systems with Applications to WASP-12}",
      journal = {\apj},
         year = 2010,
        month = oct,
       volume = {721},
       number = {2},
        pages = {923-928},
          doi = {10.1088/0004-637X/721/2/923},
archivePrefix = {arXiv},
       eprint = {1005.4497},
 primaryClass = {astro-ph.EP},
       adsurl = {https://ui.adsabs.harvard.edu/abs/2010ApJ...721..923L}
}

@ARTICLE{Li2010,
       author = {{Li}, Shu-Lin and {Miller}, N. and {Lin}, Douglas N.~C. and {Fortney}, Jonathan J.},
        title = "{WASP-12b as a prolate, inflated and disrupting planet from tidal dissipation}",
      journal = {\nat},
         year = 2010,
        month = feb,
       volume = {463},
       number = {7284},
        pages = {1054-1056},
          doi = {10.1038/nature08715},
archivePrefix = {arXiv},
       eprint = {1002.4608},
 primaryClass = {astro-ph.EP},
       adsurl = {https://ui.adsabs.harvard.edu/abs/2010Natur.463.1054L}
}

@ARTICLE{Wei2022,
       author = {{Wei}, Xing},
        title = "{Magnetic effect on equilibrium tides and its influence on the orbital evolution of binary systems}",
      journal = {\aap},
         year = 2022,
        month = aug,
       volume = {664},
          eid = {A10},
        pages = {A10},
          doi = {10.1051/0004-6361/202243486},
archivePrefix = {arXiv},
       eprint = {2206.01387},
 primaryClass = {astro-ph.SR},
       adsurl = {https://ui.adsabs.harvard.edu/abs/2022A&A...664A..10W}
}

@software{Townsend2020,
  author       = {Townsend, R.},
  title        = {MESA SDK for Mac OS},
  version      = {v24.7.1},
  year         = {2024},
  publisher    = {Zenodo},
  doi          = {10.5281/zenodo.2669543},
  url          = {https://doi.org/10.5281/zenodo.2669543}
}

@ARTICLE{Ma2021,
       author = {{Ma}, Linhao and {Fuller}, Jim},
        title = "{Orbital Decay of Short-period Exoplanets via Tidal Resonance Locking}",
      journal = {\apj},
         year = 2021,
        month = sep,
       volume = {918},
       number = {1},
          eid = {16},
        pages = {16},
          doi = {10.3847/1538-4357/ac088e},
archivePrefix = {arXiv},
       eprint = {2105.09335},
 primaryClass = {astro-ph.EP},
       adsurl = {https://ui.adsabs.harvard.edu/abs/2021ApJ...918...16M}
}

@ARTICLE{Penev2009,
       author = {{Penev}, Kaloyan and {Barranco}, Joseph and {Sasselov}, Dimitar},
        title = "{Direct Calculation of the Turbulent Dissipation Efficiency in Anelastic Convection}",
      journal = {\apj},
         year = 2009,
        month = nov,
       volume = {705},
       number = {1},
        pages = {285-297},
          doi = {10.1088/0004-637X/705/1/285},
archivePrefix = {arXiv},
       eprint = {0810.5370},
 primaryClass = {astro-ph},
       adsurl = {https://ui.adsabs.harvard.edu/abs/2009ApJ...705..285P}
}

@ARTICLE{Ogilvie2012,
       author = {{Ogilvie}, Gordon I. and {Lesur}, Geoffroy},
        title = "{On the interaction between tides and convection}",
      journal = {\mnras},
         year = 2012,
        month = may,
       volume = {422},
       number = {3},
        pages = {1975-1987},
          doi = {10.1111/j.1365-2966.2012.20630.x},
archivePrefix = {arXiv},
       eprint = {1201.5020},
 primaryClass = {astro-ph.SR},
       adsurl = {https://ui.adsabs.harvard.edu/abs/2012MNRAS.422.1975O}
}

@ARTICLE{Duguid2020b,
       author = {{Duguid}, Craig D. and {Barker}, Adrian J. and {Jones}, C.~A.},
        title = "{Convective turbulent viscosity acting on equilibrium tidal flows: new frequency scaling of the effective viscosity}",
      journal = {\mnras},
         year = 2020,
        month = sep,
       volume = {497},
       number = {3},
        pages = {3400-3417},
          doi = {10.1093/mnras/staa2216},
archivePrefix = {arXiv},
       eprint = {2007.12624},
 primaryClass = {astro-ph.EP},
       adsurl = {https://ui.adsabs.harvard.edu/abs/2020MNRAS.497.3400D}
}

@ARTICLE{Duguid2020a,
       author = {{Duguid}, Craig D. and {Barker}, Adrian J. and {Jones}, C.~A.},
        title = "{Tidal flows with convection: frequency dependence of the effective viscosity and evidence for antidissipation}",
      journal = {\mnras},
         year = 2020,
        month = jan,
       volume = {491},
       number = {1},
        pages = {923-943},
          doi = {10.1093/mnras/stz2899},
archivePrefix = {arXiv},
       eprint = {1910.06034},
 primaryClass = {astro-ph.SR},
       adsurl = {https://ui.adsabs.harvard.edu/abs/2020MNRAS.491..923D}
}

@ARTICLE{deVries2023,
       author = {{de Vries}, Nils B. and {Barker}, Adrian J. and {Hollerbach}, Rainer},
        title = "{Tidal dissipation due to the elliptical instability and turbulent viscosity in convection zones in rotating giant planets and stars}",
      journal = {\mnras},
         year = 2023,
        month = sep,
       volume = {524},
       number = {2},
        pages = {2661-2683},
          doi = {10.1093/mnras/stad1990},
archivePrefix = {arXiv},
       eprint = {2306.17622},
 primaryClass = {astro-ph.EP},
       adsurl = {https://ui.adsabs.harvard.edu/abs/2023MNRAS.524.2661D}
}

@ARTICLE{Vidal2020b,
       author = {{Vidal}, J{\'e}r{\'e}mie and {Barker}, Adrian J.},
        title = "{Efficiency of tidal dissipation in slowly rotating fully convective stars or planets}",
      journal = {\mnras},
         year = 2020,
        month = oct,
       volume = {497},
       number = {4},
        pages = {4472-4485},
          doi = {10.1093/mnras/staa2239},
archivePrefix = {arXiv},
       eprint = {2007.13392},
 primaryClass = {astro-ph.SR},
       adsurl = {https://ui.adsabs.harvard.edu/abs/2020MNRAS.497.4472V}
}

@ARTICLE{Vidal2020a,
       author = {{Vidal}, J{\'e}r{\'e}mie and {Barker}, Adrian J.},
        title = "{Turbulent Viscosity Acting on the Equilibrium Tidal Flow in Convective Stars}",
      journal = {\apjl},
         year = 2020,
        month = jan,
       volume = {888},
       number = {2},
          eid = {L31},
        pages = {L31},
          doi = {10.3847/2041-8213/ab6219},
archivePrefix = {arXiv},
       eprint = {1912.07910},
 primaryClass = {astro-ph.SR},
       adsurl = {https://ui.adsabs.harvard.edu/abs/2020ApJ...888L..31V}
}

@ARTICLE{Barker2025,
       author = {{Barker}, Adrian J.},
        title = "{Tidal interactions in stellar and planetary systems}",
      journal = {arXiv e-prints},
         year = 2025,
        month = apr,
          eid = {arXiv:2504.10941},
        pages = {arXiv:2504.10941},
          doi = {10.48550/arXiv.2504.10941},
archivePrefix = {arXiv},
       eprint = {2504.10941},
 primaryClass = {astro-ph.EP},
       adsurl = {https://ui.adsabs.harvard.edu/abs/2025arXiv250410941B}
}

@ARTICLE{Bunting2019,
       author = {{Bunting}, Andrew and {Papaloizou}, John C.~B. and {Terquem}, Caroline},
        title = "{Non-adiabatic tidal oscillations induced by a planetary companion}",
      journal = {\mnras},
         year = 2019,
        month = dec,
       volume = {490},
       number = {2},
        pages = {1784-1802},
          doi = {10.1093/mnras/stz2561},
archivePrefix = {arXiv},
       eprint = {1909.08476},
 primaryClass = {astro-ph.EP},
       adsurl = {https://ui.adsabs.harvard.edu/abs/2019MNRAS.490.1784B}
}
\bibliographystyle{aasjournal}
\end{CJK*}
\end{document}